\documentstyle[epsfile,plotfile,twoside,amsmath,amssymb,submitp]{pio}

\begin{document}

\title{Optical Solitons in Periodic Media
with Resonant and Off-Resonant Nonlinearities}
\shorttitle{Optical Solitons in  Periodic Media}

\author{
Gershon~Kurizki\thanks{E-mail: Gershon.Kurizki@weizmann.ac.il}
\and
Alexander~E.~Kozhekin\thanks{Present address: {\it Institute of
Physics and Astronomy, University of Aarhus, Ny Munkegade, DK-8000
Aarhus C, Denmark}}
\and
Tom{\'a}{\v s}~Opatrn{\'y}\thanks{Present address: {\it
Friedrich-Schiller-Universit\"at Jena, Theoretisch-Physikalisches
Institut, Max-Wien Platz 1, 07743 Jena, Germany} and {\it
Palack\'y University, Faculty of Natural Sciences, Svobody 26,
77146 Olomouc, Czech Republic} } \\
{\it  Department of Chemical Physics, \\
Weizmann Institute of Science, \\
Rehovot 76100, ISRAEL }
\and
Boris~Malomed \\
{\it
Department of Interdisciplinary Studies, \\
Faculty of Engineering, \\
Tel Aviv University, Tel Aviv 69978, Israel}
}

\maketitle

\tableofcontents
\section{Introduction}\label{S-intro}

The study of light-matter interactions in dielectric structures with
periodic modulation of the refractive index has developed into a vast
research area. At the heart of this area is the interplay between
Bragg reflections, which block the propagation of light in spectral
bands known as photonic band gaps (PBGs), and the dynamical
modifications of these reflections by {\em nonlinear\/} light-matter
interactions (see bibliography compiled by \cite{pbg-bib}).  Three- or
two-dimensional (3D or 2D) PBGs are needed in order to extinguish
spontaneous emission in all possible directions of propagation, which
requires the nontrivial fabrication of 3D- or 2D-periodic photonic
crystals (\cite{Yabl87,Yabl93}).  For controlling strictly
unidirectional propagation, it is sufficient to resort to PBGs in
one-dimensional (1D) periodic structures (Bragg reflectors or
dielectric multi-layer mirrors). Illumination of the periodic
dielectric structure at a PBG frequency in the limit of vanishing
nonlinearity leads to exponential decay of the incident field
amplitude with penetration depth, at the expense of exponential growth
of the back-scattered (Bragg-reflected) amplitude.  However, this
reflection may weaken or cease altogether, rendering the structure
transparent, when the illumination intensity and the resulting
nonlinearity modify the refractive index so as to shift (or even close
down) the PBG.  The pulsed mode of propagation in nonlinear periodic
structures exhibits a variety of fundamentally unique and
technologically interesting regimes: nonlinear filtering, switching,
and distributed-feedback amplification (\cite{Scal94,Scal94a}).  Among
these regimes, we have chosen here to concentrate on the intriguing
solitary waves existing in PBGs, known as {\em gap solitons\/} (GS),
and solitons propagating near PBGs.

A GS is usually understood as a self-localized moving or standing
(quiescent) bright region, where light is confined by Bragg
reflections against a dark background.  The soliton spectrum is tuned
away from the Bragg resonance by the nonlinearity at sufficiently high
field intensities.  There is also considerable physical interest in
finding in the vicinity of a PBG a dark soliton (DS), i.e., a `hole'
of a fixed shape in a continuous-wave (cw) background field of
constant intensity (\cite{Kivs98}).

The first type of GS was predicted to exist in a Bragg grating filled
with a Kerr medium, whose nonlinearity is cubic
(\cite{Chri89,Acev89,Feng93}).  Detailed theoretical studies of these
{\it Bragg-grating (Kerr-nonlinear) solitons} (see \cite{Ster94} for a
review) were followed by their experimental observation
(\cite{Eggl96}) in a short ($<10$ cm) piece of an optical fiber with a
resonant Bragg grating written on it.  In theoretical considerations
of solitons in a Bragg grating combined with Kerr nonlinearity, a
formidable problem is their stability. Current experiments are
conducted in short fiber pieces, which do not allow us to test the
solitons' stability.  Approximate and rigorous treatments of the
stability problem (\cite{Malo94,Bara98,DeRo98}) have demonstrated that
the Kerr-nonlinear Bragg-grating solitons have instability regions in
their frequency-amplitude parametric plane, far from the PBG-center,
small-amplitude limit. The generation of a slow gap soliton in a Kerr
- nonlinear periodic grating via Raman transfer of energy from the pump pulse
to the Bragg - resonant Stokes components was recently proposed by
\cite{Winf00}.

A Bragg grating with quadratic or second-harmonic-generating (SHG)
nonlinearity can also give rise to a rich spectrum of solitons
(\cite{Pesc97,He97,Cont97}).  This model has been shown to possess a
number of remarkable features. In particular (\cite{Pesc97}), contrary
to Kerr-nonlinear Bragg gratings, in the case of SHG nonlinearity the
gap in which solitons may exist is partly {\em empty}, only part of it being
filled with actually existing soliton solutions.

While the above temporal-domain models pertain to light propagation in
fibers, distributions of stationary fields in a planar (2D) nonlinear
optical waveguide are governed by spatial-domain equations
(\cite{Steg99}). In the planar waveguide, a Bragg grating can be
realized as a system of parallel `scratches'.  The soliton spectrum of
this model contains (\cite{Mak98}) not only the fundamental
single-humped solitons but also their two-humped bound states, which,
quite unusually, turn out to be dynamically stable. Moreover, it
possesses (\cite{Cham00}), besides the conventional GSs, also {\it
embedded solitons}, which are isolated solitary-wave solutions within
the continuous spectrum, rather than inside the gap.

A principally different mechanism of GS formation has been discovered
in a periodic array of thin layers of {\em resonant two-level systems}
(TLS) separated by half-wavelength nonabsorbing dielectric layers,
i.e., a {\em resonantly absorbing Bragg reflector} (RABR)
(\cite{Kozh95,Kozh98a,Opat99}). The RABR has been found to allow, for
{\em any} Bragg reflectivity, a vast family of stable solitons, both
standing and moving (\cite{Kozh95,Kozh98a}).  As opposed to the
$2\pi$-solitons arising in self-induced transparency, i.e., resonant
field -- TLS interaction in uniform media (\cite{McCa69,McCa70}), gap
solitons in RABR may have an {\em arbitrary} pulse area
(\cite{Kozh95,Kozh98a}).  As shown below, GS solutions have been
consistently obtained only in a RABR with {\em thin} active TLS
layers. By contrast, an attempt (\cite{Akoz98}) to obtain such
solutions in a periodic structure {\em uniformly} filled with active
TLS is physically unfounded, and fails for many parameter values, so
that such generalization remains an open problem 
(see Sec. \ref{S-App-average-fin} below). 

An unexpected property of the RABR is that, alongside the stable
bright-soliton solutions (\cite{Kozh95,Kozh98a}), this system gives
rise to a family of dark solitons (DSs), a large part of which are
{\em stable} (\cite{Opat99}).  While the existence of stable
bright-soliton solutions along with {\em un}stable DSs is a known
feature of uniform SHG media (\cite{Kivs98}), a RABR with thin active
layers provides, to the best of our knowledge, the first example of a
nonlinear optical medium in which {\em stable} bright and dark
solitons exist for the {\em same values} of the model's parameters (albeit at
different frequencies). It is also the first example of the existence
of stable bright solitons alongside {\em stable} cw ({\em background})
solutions.

Potential applications of GSs are based on the system's ability to
filter out (by means of Bragg reflections) all pulses except for those
satisfying the GS dispersion condition, as well as to control the
pulses' shape and velocity. It would be clearly desirable to
supplement these advantageous properties by immunity to transverse
diffraction of the pulse, i.e., to achieve {\em simultaneous}
transverse and longitudinal self-localization of light in a RABR. This
motivates a quest for multi-dimensional solitons that are localized in
both space and time.

The concept of  optical 
multi-dimensional spatio-temporal solitons, alias
`light bullets' (LBs), was pioneered by Silberberg (\cite{Silb90}),
and has since been investigated in various nonlinear optical
media, with particular emphasis on  the stability of LBs. 
For a SHG medium,
the existence of stable two- and three-dimensional (2D and 3D)
solitons was predicted as early as in 1981 (\cite{Kana81}), followed
by  detailed
studies of their propagation and stability against collapse
(\cite{Haya93,Malo97,Miha98,He98}), 
Recently, the first
experimental observation of a quasi-2D 
spatiotemporal soliton in a 3D SHG sample was
reported (\cite{Liu99}). 
The concept of {\em dark} LBs was 
proposed  by \cite{Chen95}). 
Stable  {\it
antidark} LBs, i.e., those supported by a finite cw background,
were predicted in a generalized nonlinear Schr\"odinger equation
which contains third-order temporal dispersion  
(\cite{Fran98}). 

As early as in 1984, simulations indicated that self-focusing of
spatiotemporal pulses in a SIT medium could result in the formation of
a quasi-stable vibrating object (\cite{Drum84}), which was a hint at
the possible existence of LBs.  It has analytically been
demonstrated (\cite{Blaa00}) that uniform 2D and 3D SIT media can
indeed carry stable LBs.  This investigation has been extended to the
case of RABR, wherein stable, transversely localized SIT solutions
combining LB and GS properties are predicted (\cite{Blaa00a}).  RABR
with {\it any }Bragg reflectivity can support stable LBs, which are
closely related to those in uniform SIT media (\cite{Blaa00}).  2D\
LBs supported by a combination of a Bragg reflector with SHG
nonlinearity were theoretically investigated still earlier
(\cite{He98}).

Finally, we briefly mention a topic that is outside the scope of this
review, namely, {\em quantum bright solitons}, which have been a
subject of extensive studies in recent years. It has been established
that solitons are superpositions of quantum states that correspond to
clusters of photons bound together.  Most of the initial activity in
this area was devoted to Kerr-nonlinear fiber solitons
(\cite{Lai89,Lai89a,Wrig91,Yuds85,Kart93,Ze91}). It was shown that
optical fiber solitons have quantum analogs which are described by the
quantum nonlinear Schr\"odinger equation (QNLSE).  The motivation was
to gain understanding of squeezing effects in soliton propagation
(\cite{Cart87,Wata89,Drum89,Haus90,Rose91} see a review by
\cite{Sizm99}), two-photon binding effects (\cite{Deut92,Deut93}), and
fundamental limits imposed on communication systems employing solitons
(\cite{Drum93,Chia93}).  Multidimensional quantum solitons were
predicted in Kerr and SHG waveguides
(\cite{Drum97,Kher98a,Kher98b,Kher00}).

The quantization of GSs has attracted considerable attention as well.
A Bethe-ansatz solution was given (\cite{Ze95}) for quantum GSs consisting
of pairwise interacting massive photons (in the effective-mass regime of PBGs
propagation) in Kerr-nonlinear 1D periodic grating. A mechanism has
been found for the creation of two-photon bound states by
photons resonantly interacting with identical two-level atoms near PBGs
in 1D periodic structures (\cite{Kuri96c}). A Bethe ansatz
solution for photons in a PBG material interacting with a single atom
was obtained by \cite{Rupa96a,Rupa96b}. It has later been generalized to an
extended many-atom periodic system, where quantum GSs involving pairs of 
photons and propagating inside a PBG have been found (\cite{John99}).
This subject is of potential interest for quantum communications via
entangled two-photon states.

This review starts with a brief survey of solitons in Bragg gratings
with cubic and quadratic nonlinearities (Ch. \ref{S-Boris}) and of
self-induced transparency (SIT) in uniform media and thin films
(Ch. \ref{sec:3}). It then continues with the derivation of the model
for phenomena similar to 
SIT in resonantly absorbing Bragg reflectors (RABR)
(Ch. \ref{S:model}).  Bright and dark solitons in RABR are discussed
in Ch. \ref{S-bright} and \ref{S-dark}, consecutively.  Light bullets
in periodic resonantly-absorbing media are treated in
Ch. \ref{sec:lightbullets}. Finally, the prospects for experimental 
progress in this area are summarized in
Ch. \ref{S-exper}. 

\section[cubic and quadratic nonlinearities]{Solitons in Bragg 
gratings with cubic and quadratic nonlinearities} 
\label{S-Boris}


\subsection{Kerr nonlinearity}

Because the Bragg grating gives rise to very strong effective
dispersion, its combination with various optical nonlinearities can
create a rich variety of solitons, which, in most cases, are {\em gap
solitons} (GSs), as their intrinsic frequency must belong to a gap in the
spectrum of  {\em linear\/} waves in the Bragg grating. Initially,
a Bragg grating filled with a Kerr nonlinear medium, whose
nonlinearity is cubic, was considered by \cite{Chri89,Acev89}, and
\cite{Feng93}. The corresponding system of two propagation equations
for the right (forward)- and left (backward)-traveling field envelopes
${\cal E}_{F}(\zeta,\tau)$ and ${\cal E}_{B}(\zeta,\tau)$  with the
self- and cross-phase-modulation cubic terms is a generalization of
the known Thirring model (in the Thirring model proper, the self-phase
modulation terms are omitted):
\begin{subequations}
\label{CompletThirring}
\begin{eqnarray}
i\frac{\partial {\cal E}_{F}}{\partial \tau}+i\frac{\partial {\cal E}_{F}}{
\partial \zeta}
+\left( |{\cal E}_{B}|^{2}+\frac{1}{2}|{\cal E}_{F}|^{2}\right) 
{\cal E}_{F}+{\cal E}_{B} &=&0,  \label{uThirring} \\
i\frac{\partial {\cal E}_{B}}{\partial \tau}-i\frac{\partial {\cal E}_{B}}{
\partial \zeta}
+\left( |{\cal E}_{F}|^{2}+\frac{1}{2}|{\cal E}_{B}|^{2}\right) 
{\cal E}_{B}+{\cal E}_{F} &=&0.  \label{vThirring}
\end{eqnarray}
\end{subequations}
Here the dimensionless time $\tau$ and length $\zeta$
are related to the physical time $t$ and length $z$ as $t$ $=$ 
$4\tau/(a_1 \omega_c)$, and  $z$  $=$ $4c\zeta/(a_1 \omega_c n_0)$, where
$n_0$ is the mean linear index of refraction,
$a_1$ is its modulation 
(see Eq.~(\ref{be1}) below), and 
$\omega_c$ is the central frequency of the band gap.
Actual values of the electric fields can be obtained from the
dimensionless quantities ${\cal E}_{F,B}$ upon multiplication by
$\omega_c \sqrt{a_1 n_0/(48\pi \chi^{(3)})}$, where
$\chi^{(3)}$ is the third order susceptibility.
In Ch.~\ref{S:model-Maxwell} we present a general derivation
of the Maxwell equations in Bragg gratings.

The existence of solitons in a given model is usually closely related
to the modulational instability of a continuous-wave (cw) solution in
the model (\cite{Agra95}): an array (chain) of bright solitons can be
generated from the cw solution by modulational instability.  In an
optical fiber combining a Bragg grating and Kerr nonlinearity, {\em
all\/} the cw states are modulationally unstable, as observed in a
direct experiment (which also involved the polarization of light) by
\cite{Slus00}.  However, as was shown by \cite{Opat99} (see
Ch. \ref{S-dark} below) a RABR gives rise to stable bright solitons
coexisting with {\em stable} cw states, which is a very unusual
property.

Unlike the Thirring model proper, its optical version based on
Eqs. (\ref{uThirring}) and (\ref{vThirring}) is not integrable.
Moreover, the equations lack any invariance with respect to 
the reference frame (i.e., the model is neither Galilean nor
Lorentz invariant).  Nevertheless, {\em exact} single-soliton
solutions, which contain two independent parameters, viz., an
intrinsic frequency $\omega $ and the soliton's velocity $v$, were
found in the above-mentioned works by \cite{Chri89} and \cite{Acev89},
following the pattern of the Thirring-model solitons. At $v=0$, the
quiescent (standing) solitons {\em completely} fill the gap $-1<\omega
<+1$ in the spectrum of the linearized equations (\ref{uThirring}) and
(\ref{vThirring}), and the velocity takes all the possible values
$-1<v<+1$ (recall that, in the present notation, the maximum group
velocity of light is $1$).  Solitons with small frequencies have
small amplitudes and are close to the classical nonlinear
Schr\"{o}dinger (NLS) solitons, while the ones with values of $\omega
^{2}$ closer to $1$ are strongly different from their NLS
counterparts; in particular, they are {\it chirped}, i.e., they have a
nontrivial intrinsic phase structure.

\input #boris1

Many properties of these {\it Bragg-grating solitons}, as they are
frequently called, were reviewed by \cite{Ster94} (see also the special
issue of {\em Optics Express\/} edited by \cite{pbg98}). Detailed
theoretical studies were finally followed by their experimental
observation by \cite{Eggl96} in a short ($<10$ cm) piece of an optical
fiber with a resonant Bragg grating written on it. This experiment
requires the use of very powerful light beams, with an intensity
comparable to the fiber's breakdown threshold. Despite this
difficulty, propagating pulses with soliton-like shapes have been
detected in the experiment, see Fig. \ref{BraggGratingSoliton}.

It should be stressed that the above-mentioned exact soliton solutions
to Eqs. (\ref{uThirring}) and (\ref{vThirring}) take the simplest form
in the case  $v=0$, 
corresponding to a pulse of {\em
standing light}. In reality, the soliton observed by \cite{Eggl96}
was moving at a considerable velocity. Generation and detection of
{\em zero-velocity} solitons in
nonlinear  Bragg gratings
remains an experimental  challenge.

Theoretical considerations of  solitons in  Bragg gratings
with  Kerr nonlinearity face the tough problem of soliton
stability. The first approach to the stability problem, developed by
\cite{Malo94} was based on the variational approximation. Although
this approach is not rigorous, it  clearly demonstrates that,
sufficiently far from the above-mentioned NLS (small-frequency
small-amplitude) limit,  Bragg-grating solitons can be {\em
un\/}stable. Later, a rigorous analysis of the same stability problem
was developed by \cite{Bara98} and by \cite{DeRo98}. It has been
demonstrated that the Bragg-grating solitons indeed have instability
regions in their parametric ($\omega , v$) plane, which are
close to those originally predicted by means of the variational
approximation.

The derivation of the standard equations (\ref{uThirring}) and
(\ref{vThirring}) from the underlying Maxwell's equations 
(see Ch..~\ref{S:model-Maxwell}) neglects all
the linear terms with second-order derivatives (which account for the
material dispersion and/or diffraction in the medium). As  shown
by \cite{Cham98}, taking the second derivatives into account
drastically changes the soliton {\em  content of the model} 
(although the propagation distance necessary to observe the change of
the solitons' shape may be much larger than that available in current
experiments). The relatively simple soliton solutions generated by
Eqs. (\ref{uThirring}) and (\ref{vThirring}) immediately disappear
after the inclusion of the second-derivative terms; instead, three
completely new branches of soliton solutions emerge, provided that the
coefficient in front of the second-derivative terms exceeds a certain
minimum value. Very recently, it has been shown by \cite{Scho00} that
these new branches are, generally, also subject to an instability.

Finally, we mention a {\em dual-core} nonlinear optical fiber, each
core containing a Bragg grating, which is described by four
equations (\cite{Mak98a}), 
\begin{subequations}
\begin{eqnarray}
i\frac{\partial  {\cal E}_{F1}}{\partial \tau}
+i\frac{\partial
{\cal E}_{F1}}{\partial \zeta} &+&
\left( \frac{1}{2}|{\cal 
E}_{F1}|^{2}+| {\cal E}_{B1}|^{2}\right)  
{\cal E}_{F1} \nonumber \\
&+& {\cal E}_{B1}+\lambda  
{\cal E}_{F2}=0,  \label{bu1} 
\\
i\frac{\partial {\cal E}_{B1}}{\partial \tau}- i\frac{\partial
{\cal E}_{B1}}{\partial \zeta} &+& \left( \frac{1}{2}| {\cal 
E}_{B1}|^{2}+| {\cal E}_{F1}|^{2}\right)  
{\cal E}_{B1} \nonumber \\
&+& {\cal E}_{F1}+\lambda 
{\cal E}_{B2}=0,  \label{bv1}
\\
i\frac{\partial  {\cal E}_{F2}}{\partial \tau}+i\frac{\partial
{\cal E}_{F2}}{\partial \zeta} &+& \left( \frac{1}{2}| {\cal 
E}_{F2}|^{2}+| {\cal E}_{B2}|^{2}\right)  
{\cal E}_{F2} \nonumber \\
&+& {\cal E}_{B2}+\lambda 
{\cal E}_{F1}=0,  \label{bu2}
\\
i\frac{\partial  {\cal E}_{B2}}{\partial \tau}- i\frac{\partial
{\cal E}_{B2}}{\partial \zeta} &+& \left( \frac{1}{2}|{\cal 
E}_{B2}|^{2}+| {\cal E}_{F2}|^{2}\right)  
{\cal E}_{B2} \nonumber \\
&+&{\cal E}_{F2}+\lambda 
{\cal E}_{B1}=0,  \label{bv2}
\end{eqnarray}
\end{subequations}
where the subscripts $1$ and $2$ stand for the core number, 
the dimensionless quantities $\tau$, $\zeta$, ${\cal E}_{B,F;1,2}$
have the same meaning as in Eq.~(\ref{CompletThirring}), and
$\lambda $ is the real linear coupling between the
cores. Due to this linear coupling, a soliton in this system
necessarily has its components in both cores. As shown by
\cite{Mak98a}, an obviously symmetric soliton with equal components in
the two cores, ${\cal E}_{F1}={\cal
E}_{F2}$ and ${\cal E}_{B1}=
{\cal E}_{B2}$, becomes unstable when its energy exceeds a
certain threshold. The instability gives rise to a {\it pitchfork
bifurcation} that generates a pair of stable asymmetric solitons which
are mirror images of each other.

\subsection{Quadratic nonlinearities}

A Bragg grating with  quadratic
(second-harmonic-generating,  SHG) nonlinearity gives rise to a vast
variety of solitons. This model was introduced
independently and simultaneously by \cite{Pesc97,He97}, and
\cite{Cont97}. The  model includes four fields, namely,
the fundamental- and second-harmonic components of the forward- (F-) and
backward- (B-) traveling waves. In a normalized notation, the corresponding 
system takes the form
\begin{subequations}
\label{quadrUV12}
\begin{eqnarray}
i \frac{\partial {\cal E}_{F}}{\partial \tau}+i\frac{\partial {\cal E}_{F}
}{\partial \zeta} +{\cal \chi E}_{F}+{\cal E}_{F}^{\ast }{\cal G}_{F}+
{\cal E}_{B} &=&0, \label{quadrU1} \\
i \frac{1}{v}\frac{\partial {\cal G}_{F}}{\partial \tau}
 +i\frac{\partial {\cal G}_{F}
}{\partial \zeta} +{\cal \chi }^{-1}{\cal G}_{F}+{\cal E}_{F}^{2}+\varkappa 
{\cal G}_{B} &=&0,  \label{quadrV1} \\
i \frac{\partial {\cal E}_{B}}{\partial \tau}-i\frac{\partial {\cal E}_{B}
}{\partial \zeta} +{\cal \chi E}_{B}+{\cal E}_{B}^{\ast }{\cal G}_{B}+
{\cal E}_{F} &=&0,  \label{quadrU2} \\
i \frac{1}{v}\frac{\partial {\cal G}_{B}}{\partial \tau}
 -i\frac{\partial {\cal G}_{B}
}{\partial \zeta} +{\cal \chi }^{-1}{\cal G}_{B}+{\cal E}_{B}^{2}+\varkappa 
{\cal G}_{F} &=&0,  \label{quadrV2}
\end{eqnarray}
\end{subequations}
where ${\cal E}$ and ${\cal G}$ are the
fundamental- and second-harmonic fields,
$v>0$ is the group velocity of the second harmonic relative
to the fundamental harmonic, $\varkappa$ is a (generally complex) coupling
F-B coefficient in the second harmonic, given that the coupling
constant at the fundamental harmonic is normalized to $1$, and the
real parameter ${\cal \chi }$ determines the {\it phase
mismatch} of the two harmonics.

Analytical and numerical consideration of solitons in this model has
revealed a number of nontrivial features. In particular, \cite{Pesc97}
have found that, contrary to the model (\ref{CompletThirring}) 
with Kerr nonlinearity, in the  
system (\ref{quadrUV12})
the gap in which solitons may exist has {\em voids}, so
that only a part of it is filled with actually existing soliton
solutions. Another noteworthy finding of the same work is that not
only single-humped fundamental solitons, but also their two-humped
bound states, which are normally unstable in other versions of the SHG
models, are {\em dynamically stable} in the four-wave model
(\ref{quadrUV12}). An example of a two-humped soliton
which turns out to be fairly stable in simulations of the system  
(\ref{quadrUV12}) is shown in
Fig. \ref{FigStable2humped4waves}. A detailed description of gap
solitons in the four-wave model combining a Bragg grating and SHG
nonlinearity can be found in the review by \cite{Etri00} on solitons in
various SHG media.

\input #boris2

A different model based on  resonant Bragg
reflection and SHG nonlinearity can be formulated in the {\em spatial
domain\/} [unlike the time-domain models 
(\ref{CompletThirring}) and (\ref{quadrUV12})], 
for stationary fields in
a planar (2D) nonlinear optical waveguide.  
In such a waveguide, a 1D Bragg grating is realized as a system of
parallel scores (`scratches').  The simplest model of this type
involves three waves (\cite{Mak98}):
\begin{subequations}
\label{3W-all}
\begin{eqnarray}
i\frac{\partial {\cal E}_{1}}{\partial \zeta}+i\frac{\partial {\cal E}_{1}}{
\partial x}+{\cal E}_{2}+{\cal E}_{3}{\cal E}_{2}^{\ast }&=&0,  \label{3W1}
\\
i\frac{\partial {\cal E}_{2}}{\partial \zeta}-i\frac{\partial {\cal E}_{2}}{
\partial x}+{\cal E}_{1}+{\cal E}_{3}{\cal E}_{1}^{\ast }&=&0,  \label{3W2}
\\
2i\frac{\partial {\cal E}_{3}}{\partial \zeta}-q{\cal E}_{3}+D\frac{\partial ^{2}
{\cal E}_{3}}{\partial x^{2}}+{\cal E}_{1}{\cal E}_{2}&=&0.  \label{3W3}
\end{eqnarray}
\end{subequations}
Here $\zeta$ and $x$ are the propagation and transverse coordinates, 
respectively; the
fields ${\cal E}_{1}$ and ${\cal E}_{2}$ are two components of the
fundamental harmonic that are transformed into each other by
resonant reflections on the 1D Bragg grating, 
${\cal E}_{3}$ is the second-harmonic component, and 
$D$ is
an effective diffraction coefficient for the second harmonic. 
The wave vectors
${\bf k}_{1,2,3}$ of the three waves are related by the resonance
condition, ${\bf k}_{1}{\bf +k}_{2} {\bf =k}_{3}$,  the real
parameter $q$ accounts for a residual phase-mismatch. The
configuration corresponding to this model assumes that the second
harmonic propagates parallel to the Bragg grating (which has the form
of the above-mentioned scores). It is therefore necessary to take
into account the diffraction of this component, while for the two
fundamental harmonics the effective diffraction induced by  resonant
Bragg scattering is much stronger than normal diffraction, which
is neglected here.

The soliton spectrum of this model is fairly rich. It contains
(\cite{Mak98})  not only fundamental
single-humped solitons but also their two-humped bound states, some of
which,  as in the case of the four-wave model (\ref{quadrUV12}), 
may be dynamically stable. A rigorous
stability analysis for various solitons in the model (\ref{3W-all}), 
based on computation of eigenvalues of the corresponding
linearized equations, was performed  by
\cite{Scho00}. This analysis has shown that some of these solitons, 
although quite stable in direct dynamical simulations, are
subject to a very weak oscillatory
instability, whereas other solitons in this model 
are stable in the rigorous sense.

The three-wave model
(\ref{3W-all}) possesses (\cite{Cham00}), 
besides the traditional GSs, numerous branches of  {\it embedded
solitons:\/}  isolated solitary-wave solutions existing within
the continuous spectrum, rather than inside the gap. Solutions of this kind
appear also
when the second-derivative terms are added to the generalized Thirring
model (\ref{CompletThirring}).

Finally, the four-wave model (\ref{quadrUV12}) with quadratic
nonlinearity can be extended to the two- and three-dimensional cases,
by adding transverse diffraction terms to each equation of the
system. Physically, this generalization corresponds to spatiotemporal
evolution of the fields in a two- or three-dimensional layered medium.
Because, as is well known, quadratic nonlinearity does not give rise
to wave collapse in any number of physical dimensions, the latter
model can support stable spatiotemporal solitons, frequently called
{\it light bullets}.  Direct numerical simulations reported by
\cite{He98} have confirmed the existence of stable `bullets' in a
multidimensional SHG medium embedded in a Bragg grating.


\section[Self-Induced Transparency]{Self-Induced Transparency (SIT)
in Uniform Media and Thin Films} \label{sec:3}

\subsection{SIT in Uniform media}
\label{sit_in_um}

Self-induced transparency (SIT) is the solitary propagation of
electromagnetic (EM) pulses in near-resonant atomic media,
irrespective of the carrier-frequency detuning from resonance.  This
striking effect, which is of pa\-ra\-mount importance in nonlinear optics,
was discovered by \cite{McCa69,McCa70}.  If the pulse duration is much
shorter than the transition (spontaneous-decay) lifetime ($T_1$) and
dephasing time ($T_2$), then the leading edge of the pulse is
absorbed, inverting the atomic population, while the remainder of the
pulse causes atoms to emit stimulated light and thus return the energy
to the field. When conditions for the process are met, it is found
that a steady-state pulse envelope is established and then propagates
without attenuation at a velocity that may be considerably less than
the phase velocity of light in the medium.

We start with the Hamiltonian for a single atom in the field,
\begin{equation}
 \label{2.ham} \hat H = \frac{\hbar \omega_{0}}{2} \hat w - {\bi E}
 \cdot \hat {\bi d} ,
\end{equation}
where
\begin{equation}
 \label{2.a} \hat w \equiv |e\rangle \langle e| - |g\rangle \langle g
 | ,
\end{equation}
is the atomic inversion operator, $\omega_{0}$ is the atomic
transition frequency, $|g\rangle$ and $|e\rangle$ denote the atomic
ground and excited states, respectively, ${\bi E}$ is the electric
field vector and $\hat {\bi d}$ is the atomic dipole-moment operator.
We take the projection on the field direction, so that ${\bi E} \cdot
\hat {\bi d}$ = $E \hat d$, where
\begin{equation}
 \label{2.b} \hat d \equiv \frac{\mu}{2} \left( \hat P + \hat P^{\dag}
 \right) ,
\end{equation}
$\mu$ being the dipole moment matrix element (chosen real) and
\begin{equation}
 \label{2.c} \hat P \equiv 2\ |g\rangle \langle e| 
\end{equation}
atomic polarization operator.

We express the electric field at a given point by means of the Rabi
frequency $\Omega$ as
\begin{equation}
 \label{2.E0} E = \frac{\hbar}{2\mu} (\Omega e^{-i\omega_{c} t} +
 \Omega^{*}e^{i\omega_{c} t}) .
\end{equation}
The Heisenberg equations of motion $d \hat A/dt$ $=$ $1/(i\hbar)$
$\left[ \hat A, \hat H \right]$, for the atomic polarization and
inversion operators (\ref{2.c}),(\ref{2.a}), yield the Bloch equations
for their expectation values (c-numbers) $P$ and $w$, respectively
\begin{subequations}
\label{eq:3.1}
\begin{eqnarray}
\partial_t {P}(z,t) &=& w(z,t) \Omega - 
i (\omega_0 - \omega_{12} ) {P}, \\ 
\partial_t w(z,t) &=& -\frac{1}{2} \left[ 
{P}^*(z,t) \Omega + \mbox{ c.c. } \right].
\end{eqnarray}
\end{subequations}
The Maxwell equations (\cite{Newe92}) reduce in the rotating-wave and
slow-varying approximations to
\begin{equation}
\label{eq:3.2}
\left( \frac{c}{n_0} \frac{\partial}{\partial z} + 
\frac{\partial}{\partial t}\right) \Omega = \tau_0^{-2} {P}, 
\end{equation}
where 
\begin{equation}
\tau _0= \frac{n_0}{\mu}\sqrt{\frac{\hbar }{2\pi \omega_{c} \varrho_{0}} },  
\label{tau0}
\end{equation}
is the cooperative resonant absorption time, 
$\varrho_{0}$ being the TLS density (averaged over $z$),
and $n_0$ is the refraction index of the host media.

In the simplest case, when the driving field is in resonance with the
atomic transition, $\omega_0=\omega_{c}$, the Bloch equations
(\ref{eq:3.1}) can be easily integrated and the Maxwell equation
(\ref{eq:3.2}) then reduces to the sine-Gordon equation
\begin{equation}
\label{eq:3.3}
\frac{\partial^2 \theta}{\partial \zeta \partial \tilde{\tau}} = -\sin{\theta}
\end{equation}
for the `rotation angle',
\begin{equation}
\label{eq:3.4}
\theta=\int_{-\infty}^{t} \Omega \, d t',
\end{equation}
in terms of the dimensionless variables $\tilde{\tau}=(t-n_0 z /c)/\tau_0$ and
$\zeta=n_0 z/c\tau_0$.

This sine-Gordon equation is known to have solitary-wave solutions,
for which the total area under the pulse is conserved and equal to $2
\pi$ -- the so called pulse-area theorem by
\cite{McCa69,McCa70}:
\begin{equation}
{\Omega}(\zeta, \tilde{\tau}) = 
(\tau_o)^{-1} A_0 \mbox{\rm sech} \left[ \beta (\zeta-v \tilde{\tau})
\right] ,
\label{2pi-SG}
\end{equation}
where pulse width $\beta$ is an arbitrary real parameter uniquely
defining amplitude $A_0=2/\beta$ and group velocity $v=1/\beta^2$ of
the soliton.

Since its inception, SIT has become an active
research area with many practical applications, for which we refer
readers to excellent reviews by \cite{Lamb71,Polu75,Maim90} and
references therein. In this section we will only briefly discuss
results which are pertinent to the present review, such as SIT in thin
films and collisions of counter-propagating SIT solitons. 

\subsection{SIT in thin films}

The interaction of light with a thin film of a nonlinear resonant
medium located at the interface between two linear media has been
described by \cite{Rupa82} and \cite{Rupa87}, who have shown that a
nonlinear thin film of TLS can be a nearly ideal mirror for weak
pulses, but transparent for pulses of sufficient intensity.  The
problem of light pulse transmission through the nonlinear medium
boundary has been studied under conditions of coherent interaction
with the matter. The system can be described by a set of nonlinear
Maxwell-Bloch-like equations which effectively take the presence of
the reflected wave into account by imposing boundary conditions on the
electromagnetic fields at the interface. It has been shown
(\cite{Rupa87}) that these equations are exactly integrable by the
inverse scattering method, and $2 \pi$-soliton-pulse transmission
through the film has been studied.

If the atomic density is such that on average there are more than one
atom per cubic resonant wavelength, then near-dipole-dipole (NDD)
interactions, or local-field effects, can no longer be ignored,
contrary to the case of more dilute media. NDD effects necessitate a
correction to the field that couples to an atom in terms of the
incident field and volume polarization (\cite{Bowd91,Scal95}). This
effect can give rise to {\em bistable optical transmission} of
ultrashort light pulses through a thin layer consisting of two-level
atoms (\cite{Bash88,Bene91}): the local-field correction leads to an
inversion-dependent resonance frequency, and generates a new mechanism
of nonlinear transparency. When the excitation frequency is somewhat
larger than the original resonant frequency, the transmission of the
layer exhibits a transient bistable behavior on the time scale of
superradiance (\cite{Bash88,Bene91}). It was shown that if an
ultrashort pulse is allowed to interact with a thin film of optically
dense two-level systems, the medium response is characterized by a
rapid switching effect (\cite{Cren92,Cren92a}). This behavior is more
remarkable than the response of conventional two-level systems,
because the medium can only be found in one of two states: either
fully inverted, or in the ground state, depending (quasi-periodically)
on the ratio between the peak field-strength and the NDD coupling
strength. This feature was found to be impervious to changes in pulse
shape, and independent of the pulse area (\cite{Cren92}).

Passage of light through a system of two thin TLS films of two-level
atoms has been considered by \cite{Logv92,Babu00} who have shown that
if the distance between the films is an integer multiple of the
wavelength, then the system is bistable. Self-pulsations, i.e.,
periodically generated output, arise if an odd number of
half-wavelengths can be fitted between the films and absorption in the
medium is insignificant.  In general, the dynamics admits both regular
and chaotic regimes.

\subsection{Collisions of counterpropagating SIT solitons}\label{sec:colls}

Situations in which it is necessary to consider the interaction of
incident (forward) and reflected (backward) light waves include:
intrinsic optical bistability (\cite{Ingu90}), dynamics of excitations
in a cavity (\cite{Shaw90}) and collisions of counterpropagating SIT
solitons (\cite{Afan90,Shaw91}).  The field in such problems is
represented as a superposition of forward- and backward-traveling
waves. The atomic response to this field is determined by solving the
Bloch equations (\ref{eq:3.1}) in the rotating-wave approximation. The
population inversion $w(z,t)$ and polarization $P$ may be represented
by a quasi-Fourier expansion over a succession of spatial harmonic
carriers and slow varying envelopes, entangled in a fashion which
leads to an {\em infinite hierarchy} of equations. {\em The truncation
of this hierarchy can only be justified by phenomenological
arguments\/}, such as atom movement in an active atomic gas.

When the forward (F-) and backward (B-) wave pulses overlap in space
and time, the resulting interference pattern of nodes modifies the
atomic excitation pattern. The spatial quasi-Fourier expansion
provides an efficient way of treating the spatial inhomogeneities of
the response in those regions where the F- and B-pulses overlap, 
each successive Rabi cycle increasing the number of terms that
contribute to the expansion (\cite{Shaw90}).

Collisions of optical solitons produce observable effects on both the
atoms and the pulses. The overlap of two counterpropagating pulses can
produce an appreciable spatially-localized inversion of the atomic
population, thus causing optical solitons to lose energy. It was found
that, whereas large-energy solitons passed freely through each other,
solitons whose initial energy fell below a critical value were
destroyed by collisions. In addition, the residual atomic dipole,
created by the excitation, acts as a further source of radiation. This
radiation appears as an oscillating tail on the postcollisional pulses
and, over longer time scales, as fluorescence (\cite{Afan90,Shaw91}).



\section[SIT in RABR:  
model]{SIT in Resonantly Absorbing
Bragg Reflectors (RABR): The Model} \label{S:model}

\subsection{Maxwell equations}
\label{S:model-Maxwell}

Let us assume (\cite{Kozh95,Kozh98a,Opat99}) a one-dimensional (1D)
periodic modulation of the linear refractive index
$n(z)$ along the $z$ direction of the electromagnetic wave propagation
(see Fig. \ref{figschem}). The
modulation can be written as the Fourier series
\begin{equation}
n^2(z)=n_0^2[1+a_{1}\cos (2k_{c}z) + a_{2}\cos (4k_{c}z) + \dots],  \label{be1}
\end{equation}
where $n_0$, $a_{j}$ and $k_{c}$ are constants, and the medium is
assumed to be infinite and homogeneous in the $x$ and $y$ directions.

The periodic grating gives rise to photonic
band gaps (PBGs) in the system's linear
spectrum, i.e., the medium is totally reflective for waves whose
frequency is inside the gaps.  The central frequency of the
fundamental gap is $\omega_{c} =k_{c}c/n_0$, $c$ being the vacuum
speed of light, and the gap edges are located at the frequencies
\begin{equation}
\omega _{1,2}=\omega_{c} \left( 1\pm a_1/4\right) ,  \label{be3}
\end{equation}
where $a_1$ is the modulation depth from Eq. (\ref{be1}). 

We further
assume that {\em very thin\/} TLS layers (much thinner than $1/k_{c}$),
whose resonance frequency $\omega_0$ is close to the gap center
$\omega_{c} $, are placed at the maxima of the modulated refraction
index.  In other words, the thin active layers are placed at the
points $z_{{\rm layer}}$ such that $\cos (k_{c}z_{{\rm layer}})=\pm1$.

\input #structurescheme

We shall study the propagation of  electromagnetic waves with
frequencies close to $\omega_{c} $ through the described medium.  Let
us write the Maxwell equation for one component of the field vector
propagating in the $z$ direction as
\begin{equation}
 \label{max}
 c^2 \frac{\partial ^2 E}{\partial z^2} - n^2(z) 
 \frac{\partial ^2 E}{\partial t^2} = 
 \frac{\partial ^2 P_{nl}}{\partial t^2} ,
\end{equation}
with the refraction index $n$ modulated as in Eq.~(\ref{be1}), $E$
being the electric field component and $P_{nl}$ the non-linear
polarization.  We use the substitution
\begin{equation}
 \label{1.esubst}
 E \equiv \left[ {\cal E}_{F}(z,t) e^{ik_{c}z} +
{\cal E}_{B}(z,t) e^{-ik_{c}z}  \right] e^{-i \omega_{c} t} + {\rm c.c.},
\end{equation}
with $\omega_{c}$ satisfying the dispersion relation $n_0 \omega_{c} =
k_{c}c$ and ${\cal E}_{F}$ and ${\cal E}_{B}$ denoting the forward and
backward propagating field components. We work in the
slowly-varying envelope approximation 
\begin{subequations}
\begin{eqnarray}
 \label{1.smallnes}
 \left| \frac{\partial ^2 {\cal E}_{B,F}}{\partial z^2} \right| 
 & \ll & \left| k_{c} \frac{\partial  {\cal E}_{B,F}}{\partial z} \right| , \\
 \label{2.smallnes}
 \left| \frac{\partial ^2 {\cal E}_{B,F}}{\partial t^2} \right| 
 & \ll & \left| \omega_{c} \frac{\partial  {\cal E}_{B,F}}{\partial t} \right| . 
\end{eqnarray}
\end{subequations}
Substituting (\ref{1.esubst}) into (\ref{max}), using
(\ref{1.smallnes}) and (\ref{2.smallnes}), multiplying by $e^{i(\mp
k_{c}z+\omega_{c} t)}$ and averaging over the wavelength $\lambda$ $=$
$2\pi/k_{c}$ and the period $T$ $=$ $2\pi /\omega_{c}$ we get
\begin{subequations}
\label{Maxw-P}
\begin{eqnarray}
 \frac{c}{n_0} \frac{\partial  {\cal E}_{F}}{\partial z} 
 + \frac{\partial  {\cal E}_{F}}{\partial t}
 & = & \frac{i a_{1} \omega_{c}}{4} {\cal E}_{B} 
+ \frac{\hbar}{2\mu \tau_{0}^{2}} P_{-} ,
 \label{1.b1}
 \\
  -\frac{c}{n_0} \frac{\partial  {\cal E}_{B}}{\partial z} 
 + \frac{\partial  {\cal E}_{B}}{\partial t}
& = & \frac{i a_{1} \omega_{c}}{4} {\cal E}_{F} 
+ \frac{\hbar}{2\mu \tau_{0}^{2}} P_{+} ,
 \label{1.b2} 
\end{eqnarray}
\end{subequations}
where we define
\begin{equation}
 P_{\pm} \equiv -\frac{i\mu \tau_{0}^{2}}{\hbar \omega_{c}n_{0}^{2}} 
 \left\langle
 \frac{\partial ^2 P_{nl}}{\partial t^2} e^{i(\pm k_{c}z+\omega_{c} t)}
 \right\rangle _{\lambda,T} ,
 \label{blabol1}
\end{equation}
with $\mu$ being the dipole moment (Ch.\ref{sec:3}, and $\tau_{0}$ 
a constant chosen here to be the medium absorption time 
[see Eq.~(\ref{tau0})].
The averaging is  defined as
\begin{equation}
 \label{1.c}
 \big\langle \dots
 \big\rangle _{\lambda,T} \equiv
 \frac{1}{\lambda T}\int_{\lambda} \int_{T} \dots dt \ dz .
\end{equation}
We express the field components ${\cal E}_{F,B}$ by means of the
dimensionless quantities $\Sigma_{\pm}$,
\begin{equation}
 {\cal E}_{F,B} = \frac{\hbar}{4\mu \tau_{0}} \left( \Sigma_{+} \pm
  \Sigma_{-}  \right) ,
  \label{calesigma}
\end{equation}
so that the electric field $E$ $=$ $E(z,t)$ is
\begin{eqnarray}
E(z,t)=\hbar (\mu \tau _0)^{-1}\left( {\rm Re}\ \left[ \Sigma
_{+}(z,t)e^{-i\omega_{c} t}\right] \cos k_{c}z  
\right. \nonumber 
\\ 
\left.
-{\rm Im}\ \left[ \Sigma
_{-}(z,t)e^{-i\omega_{c} t}\right] \sin k_{c}z\right) .  \label{be4}
\end{eqnarray}

To obtain the equations of motion in the most compact form, it is
convenient to introduce the dimensionless time $\tau $, coordinate
$\zeta $, and detuning $\delta $ as
\begin{equation}
  \tau \equiv t/\tau _0,\;\zeta \equiv \left( n_0/c\tau _0\right) z,\;\delta
  \equiv (\omega_0-\omega_{c} )\tau _0, 
  \label{dimensls}
\end{equation}
where $\tau _0$ is the {\em characteristic absorption time\/} of the
field by the TLS medium as defined by Eq.~(\ref{tau0}).
Substituting Eq.~(\ref{calesigma}) into (\ref{1.b1}) and (\ref{1.b2}),
using (\ref{dimensls}) and (\ref{tau0}), then
differentiating the equations again with respect to $\zeta$ and $\tau$,
we arrive, after algebra, at the form
\begin{subequations}
\begin{eqnarray}
 \frac{\partial ^2\Sigma _{+}}{\partial \tau ^2}-
 \frac{\partial ^2\Sigma _{+}}
 {\partial \zeta ^2} =-\eta ^2\Sigma _{+}+i\eta (P_{+}+P_{-})
 ,\nonumber \\
 + \frac{\partial }
 {\partial \tau } \left( P_{+}+P_{-} \right) +
 \frac{\partial }{\partial \zeta } \left( P_{+}-P_{-} \right),
 \label{Sigplap}
\end{eqnarray}
and
\begin{eqnarray}
 \frac{\partial ^2\Sigma _{-}}{\partial \tau ^2}-
 \frac{\partial ^2\Sigma _{-}}
 {\partial \zeta ^2} =-\eta ^2\Sigma _{-}+i\eta (P_{+}-P_{-}) \nonumber \\
 - \frac{\partial }
 {\partial \tau } \left( P_{+}-P_{-} \right) -
 \frac{\partial }{\partial \zeta } \left( P_{+}+P_{-} \right).
 \label{Sigmiap}
\end{eqnarray}
\end{subequations}
Here the dimensionless modulation strength $\eta$ is the ratio of the
TLS {\em absorption distance\/} $l_{\rm abs}=\tau_0 c/n_0$ to the {\em
Bragg reflection distance\/} $l_{\rm refl}=4c/(a_1 \omega_c n_0)$,
which can be expressed as
\begin{equation}
 \label{eqeta}
 \eta =l_{\rm abs}/l_{\rm refl} = a_{1}\omega_{c} \tau _0/4.
\end{equation} 

The equations for the electric field components $\Sigma _{\pm}$ can be
solved once we know the averaged polarization $P_{\pm}$.  To find $P_{\pm}$, we express the polarization $P_{nl}$ at a
given point as the dipole moment density
\begin{eqnarray}
 \label{3.1}
 P_{nl} &=& 4\pi \varrho \langle d \rangle = 2 \pi \varrho \mu \left( 
 \langle \hat P \rangle  + \langle \hat P^{\dag} \rangle \right)
 \nonumber \\
 &=& - 2\pi i \varrho \mu \left( P e^{-i\omega_{c} t}
 - P^{*} e^{i\omega_{c} t} \right) ,
\end{eqnarray}
where $\varrho$ is the number of the two-level atoms in a unit volume.
Neglecting the time derivatives of $P$ with respect to those of
$e^{-i\omega_{c} t}$, we can write for the polarization derivative
\begin{eqnarray}
 \label{3.2}
 \frac{\partial ^{2} P_{nl}}{\partial t^{2}} = 
 2\pi i \omega_{c}^{2}\varrho \mu \left(
 P e^{-i\omega_{c} t} - P^{*} e^{i\omega_{c} t}  \right) , 
\end{eqnarray}
so that $P_{\pm}$ on the right-hand side of Eqs. (\ref{1.b1}),
(\ref{1.b2}) read as
\begin{eqnarray}
 \label{3.3}
P_{\pm} = \frac{2 \pi \omega_{c} ^{2} \mu^{2} \tau_{0}^{2}}{\hbar n_{0}^{2}}
 \left\langle \varrho P e^{\pm ik_{c}z}
 \right\rangle _{\lambda} .
 \label{blabol2}
\end{eqnarray}

To determine the evolution of $P_{\pm}$, we have to make certain
approximations.  Let us first consider the situation when the TLS's
are confined to {\em infinitely thin\/} layers. We will subsequently
study the influence of finite width of the layers.


\subsection{Two-level systems (TLS) in infinitely thin layers}

Let us now assume that the atomic density  $\varrho$ is
concentrated in zero-width layers located at $z_{j}$ such that
\begin{equation}
 e^{i k_{c}z_{2j}} = 1, \qquad e^{i k_{c}z_{2j+1}} = -1
 \label{expons}
\end{equation}
i.e., it is described by
\begin{equation}
 \label{3.4}
 \varrho = \frac{\varrho_{0} \lambda}{2} \sum_{j} \delta(z-z_{j}) ,
\end{equation}
where $\varrho_{0}$ is the bulk density averaged over the whole
wavelength.  We also assume that the  spatial
dependence of $P$ is to a
good approximation {\em anti-periodic\/} with respect to $\lambda /2$,
i.e. $P(z+\lambda/2)$ $\approx$ $-$ $P(x)$.  This is in agreement with
the approximate anti-periodicity of the electric field with $\lambda
/2$.  Denoting by $P_{0}$ the value of $P$ in even-numbered layers 
(the value in the odd-numbered layer 
being $-P_{0}$), we get the spatial
average in (\ref{3.3}) as
\begin{equation}
 \label{3.3var}
 P_{\pm} =
 \frac{2\pi \omega_{c} \mu ^{2}\varrho_{0} \tau_{0}^{2}}{\hbar n_{0}^{2}}
  P_{0}  .
\end{equation}
Note that $P_{+}$ $=$  $P_{-}$ is the consequence of the zero width
of the layers.
Due to the choice of $\tau_{0}$ as in Eq.~(\ref{tau0}), we obtain the simple
relation
\begin{equation}
 \label{3.3vara}
 P_{\pm} =  P_{0}  .
\end{equation}
Using this expression in (\ref{Sigplap}) and (\ref{Sigmiap}), we
obtain the following evolution
equations for $\Sigma _{\pm}$ 
\begin{subequations}
\label{eq:max}
\begin{eqnarray}
  \frac{\partial ^2\Sigma _{+}}{\partial \tau ^2}-
  \frac{\partial ^2\Sigma _{+}}
  {\partial \zeta ^2} &=&-\eta ^2\Sigma _{+}+2i\eta P+2\frac{\partial P}
  {\partial \tau },  \label{Sigma} \\
   \frac{\partial ^2\Sigma _{-}}{\partial \tau ^2}-\frac{\partial ^2\Sigma _{-}}
  {\partial \zeta ^2}&=&-\eta ^2\Sigma _{-}-2\frac{\partial P}{\partial \zeta
  },
  \label{driven}
\end{eqnarray}
\end{subequations}
where we have omitted the index 0 of $P_{0}$, for simplicity.

The equations for the polarization $P$ and inversion $w$ in the
even-numbered layers 
can be obtained from Eqs.~(\ref{eq:3.1}) by substituting for
$\Omega$ (combining Eqs.~(\ref{be4}) and (\ref{2.E0}))
\begin{equation}
 \Omega=\tau_{0}^{-1}\left( \Sigma_{+}\cos k_{c}z + i \Sigma_{-}
 \sin k_{c}z \right)
 \label{blabomega}
\end{equation}
and applying Eq.~(\ref{expons})  at the positions
of these layers. 
Expressing the detuning as in Eq.~(\ref{dimensls}), we
obtain the equations
\begin{subequations}
\label{eq:bloch}
\begin{eqnarray}  
  \frac{\partial P}{\partial \tau } &=&-i\delta P+\Sigma _{+}w,  \label{P} \\
  \frac{\partial w}{\partial \tau } &=&-{\rm Re}\ \left( \Sigma
  _{+}P^{*}\right) .  \label{w} 
\end{eqnarray}
\end{subequations}

The set of equations (\ref{eq:max}) and (\ref{eq:bloch}) was first
derived by \cite{Mant84} and \cite{Mant86} for the particular case of a
periodic array of thin TLS layers {\em without\/} modulation of the
linear index of refraction, i.e., $\eta=0$. In this case, these
equations can be reduced to the sine-Gordon equation (\ref{eq:3.3})
for the area of the `forward' wave
\begin{equation}
\label{eq:3.6}
\Psi(z,t)=\int_{-\infty}^{t}  \Sigma_{+}(z,t') dt'.
\end{equation}
It follows from (\ref{eq:3.6}) that the $2 \pi$ solitary wave (SIT),
associated with the sum of 
the forward and backward propagating waves, is
an {\em exact\/} solution to the coupled-mode Maxwell-Bloch equations
for a periodic stack of resonant thin films satisfying the Bragg
condition.  These solitary wave solutions are referred to as
``two-wave solitons'' (\cite{Mant86}).

If the Bragg condition is not exactly satisfied, then 
the system can
exhibit a rich, multi-stable behavior, which has been studied numerically
(\cite{Mant92,Lako92,Lako94}). In the case of a slight violation of the
Bragg condition at the
exact two-level resonance, an analytical solution
for the 
2$\pi$ gap soliton  with a phase modulation had been obtained
(\cite{Mant95}). These studies have 
also  revealed the existence of an
oscillating pulse, whose amplitude and velocity sign change
periodically.

Combining Eqs. (\ref{P}) and (\ref{w}), one can eliminate the TLS
population inversion:
\begin{equation}
w=\pm \sqrt{1-|P|^2}.  \label{eliminate}
\end{equation}
Without the field-induced polarization, the TLS population is not
inverted ($w=-1$), hence the lower sign must be chosen in
Eq. (\ref{eliminate}).  Thus, the remaining equations for $\Sigma
_{+}$ and $P$ form a {\em closed system},
\begin{subequations}
\label{eq:system}
\begin{eqnarray}
\frac{\partial ^2\Sigma _{+}}{\partial \tau ^2}-\frac{\partial ^2\Sigma _{+}%
}{\partial \zeta ^2} =&-&\eta ^2\Sigma _{+}+2i(\eta -\delta )P
\nonumber \\
~ &-& 2\sqrt{1-|P|^2%
}\ \Sigma _{+},  \label{bef1} \\
\frac{\partial P}{\partial \tau }=&-&i\delta P-\sqrt{1-|P|^2}\ \Sigma
_{+},  \label{bef2}
\end{eqnarray}
\end{subequations}
and $\Sigma _{-}$, the field component driven by
$\partial P / \partial \zeta$, can then be found from Eq. (\ref{driven}).


\subsection{Finite width of TLS layers}
\label{S-App-average-fin}

So far, we have assumed that the TLS layers are infinitesimally thin.
We now proceed to 
estimate effects of non-zero width of the active layers,
which represent more realistic physical situations.  We still
assume the width of the active layers to be small in comparison to the
wavelength. This allows us to expand the polarization as a Taylor
series in the position within the layer, and consider only terms up to
the second order. Averaging the polarization over the entire
wavelength yields the source term of the Maxwell equations (\ref{Maxw-P}).

Let us assume that the TLS density is given, instead of
Eq.~(\ref{3.4}), by
\begin{equation}
 \label{3.4b}
 \varrho = \frac{\varrho_{0} \lambda}{2} f(x) ,
\end{equation}
where $\varrho_{0}$ is the density averaged over the whole wavelength
and the function $f(z)$ has the properties
\begin{eqnarray}
 \label{3.5}
 f(z) \ge 0, \qquad f(z+\lambda/2) &=& f(z) ,
 \nonumber \\
 \int_{z_j-\frac{\lambda}{4}}^{z_j+\frac{\lambda}{4}} f(z) \ dz &=& 1,
 \nonumber \\
 \int_{z_j-\frac{\lambda}{4}}^{z_j+\frac{\lambda}{4}} z\ f(z) \ dz &=& 0,
 \nonumber \\
 \int_{z_j-\frac{\lambda}{4}}^{z_j+\frac{\lambda}{4}} z^2\ f(z) \ dz &=&
 \frac{\gamma^2}{k^2}  ,
 \label{defgamma}
\end{eqnarray}
where $\gamma \ll 1$ is a dimensionless parameter describing the
thickness of the layers. For a rectangular function $f$ of width
$D$ this parameter is
\begin{equation}
  \label{app-gam}
  \gamma = \frac{\pi D}{\sqrt{3} \lambda}\,. 
\end{equation}
To calculate $\left\langle \varrho P e^{\pm ik_{c}z}
\right\rangle _{\lambda}$ of (\ref{blabol2}), we expand the spatial dependence
of $P$ and exp$(\pm ik_{c}z)$ in Taylor series,
\begin{eqnarray}
 \label{3.8} 
 \left\langle \varrho P e^{\pm ik_{c}z}
 \right\rangle _{\lambda} =  e^{\pm ik_{c}z_{2j}} \varrho_0 
 \int_{z_{2j}-\frac{\lambda}{4}}^{z_{2j}+\frac{\lambda}{4}} 
 f(z) \nonumber \\
 \times
 \left[P_0 + P_0 '(z\!-\!z_{2j}) + \frac{1}{2}P_0 '' (z\!-\!z_{2j})^2 
 + \dots \right]
 \nonumber \\
 \times
 \left[1 \pm ik_{c}(z\!-\!z_{2j}) - \frac{k_{c}^2}{2} 
 (z\!-\!z_{2j})^2 + \dots \right]  dz,
\end{eqnarray}
where $P_0$, $P_0'$, and $P_0''$ refer to the values of $P$ and its
spatial first and second derivatives at the positions
of the even-numbered layers.  Neglecting
higher than quadratic terms and using  Eqs.
(\ref{3.5}), we arrive at
\begin{equation}
 \label{3.9} 
 \left\langle \varrho P e^{\pm ik_{c}z}
 \right\rangle _{\lambda} \approx  \varrho_0 
 \left[P_0 + \gamma^2 \left( \frac{P_0''}{2k^2} \pm \frac{iP_0'}{k}
 -\frac{P_0}{2}
 \right) \right] ,
\end{equation}
so that
\begin{equation}
 \label{3.9blab} 
 P_{\pm} \approx  
 P_0 + \gamma^2 \left( \frac{P_0''}{2k^2} \pm \frac{iP_0'}{k}
 -\frac{P_0}{2}
 \right)  .
\end{equation}
Using these values in (\ref{blabol2}), substituting them into
(\ref{Sigplap}) and (\ref{Sigmiap}), and defining
\begin{eqnarray}
  \label{app0}
  P_{0} &\equiv&  P (z_{2j}), \\
  P_{B} &\equiv& -\frac{2i}{k} \left. \frac{\partial P}{\partial z}
  \right| _{z=z_{2j}} , 
  \\
  P_{C} &\equiv& \left. 
  \left( \frac{1}{k^2}\frac{\partial ^{2}}{\partial z^{2}}
  - 1 \right) P
  \right| _{z=z_{2j}},
  \label{apps} 
\end{eqnarray}
we obtain the field equations of motion in the form 
\begin{subequations}
\begin{eqnarray}
 \label{4.17}
 \frac{\partial ^2 \Sigma_+}{\partial \tau ^2} -
 \frac{\partial ^2 \Sigma_+}{\partial \zeta ^2}
 &=& -\eta ^2 \Sigma_+  + 2i \eta P_0 
 + 2\frac{\partial P_0}{\partial \tau} 
 \nonumber \\ ~ &~&
 +
 \gamma^2
 \left[ i\eta P_{C} + \frac{\partial P_{C}}{\partial \tau}
 - \frac{\partial P_{B}}{\partial \zeta} \right] ,
 \\
 \label{4.18}
 \frac{\partial ^2 \Sigma_-}{\partial \tau ^2} -
 \frac{\partial ^2 \Sigma_-}{\partial \zeta ^2}
 &=& -\eta ^2 \Sigma_-  -2 \frac{\partial P_0}{\partial \zeta}
 \nonumber \\ ~ &~&
  + \gamma^2
 \left[ - i\eta P_{B} - \frac{\partial P_{C}}{\partial \zeta }
 - \frac{\partial P_{B}}{\partial \tau} \right] . 
\end{eqnarray}
\end{subequations}

To obtain the equations of motion for the atomic parameters, we use
Eq.~(\ref{blabomega}) for $\Omega$ in (\ref{eq:3.1}) and rewrite the
Bloch equations at a point $z$ as
\begin{subequations}
\begin{eqnarray}
 \label{5.3}
 \frac{\partial P}{\partial \tau} &=& -i\delta P
 + \left[
 \cos (k_{c}z) \Sigma_+ + i \sin (k_{c}z) \Sigma _-
 \right] w , \\
 \label{5.4}
 \frac{\partial w}{\partial \tau} &=&
 -\frac{1}{2} \left[ \cos (k_{c}z) \Sigma_+ + i \sin (k_{c}z)
 \Sigma _- \right] P^{*} + {\rm c.c.} .
\end{eqnarray}
\end{subequations}
Considering that the spatial derivatives of $\Sigma _{\pm}$
are much smaller than those of sin$(k_{c}z)$ and cos$(k_{c}z)$, and
calculating the first and second derivatives at the 
positions of the even-numbered layers, we get
\begin{eqnarray}
 \label{5.5}
 \frac{\partial P_0}{\partial \tau} &=& -i\delta P_0
 + \Sigma_+ w_0 , \\
 \label{5.6}
 \frac{\partial P_0 '}{\partial \tau} &=& -i\delta P_0 '
 + ik_{c}\Sigma_- w_0 + \Sigma_+ w_0 ' , \\
 \label{5.7}
 \frac{\partial P_0 ''}{\partial \tau} &=& -i\delta P_0 ''
 -k_{c}^2 \Sigma_+ w_0 + 2ik_{c}\Sigma_- w_0 ' + \Sigma_+ w_0 '' ,  
\end{eqnarray}
and
\begin{eqnarray}
 \label{5.8}
 \frac{\partial w_0}{\partial \tau} &=& -\frac{1}{2} \Sigma_+ P_0^{*} + {\rm
 c.c.}, \\
 \label{5.9}
 \frac{\partial w_0 '}{\partial \tau} &=& -\frac{ik_{c}}{2} \Sigma_- P_0 ^{*}
 -\frac{1}{2} \Sigma_+ P_0 ^{\prime *} + {\rm
 c.c.}, \\
 \label{5.10}
 \frac{\partial w_0 ''}{\partial \tau} &=& \frac{k_{c}^2}{2} \Sigma_+ P_0^{*}
 -ik_{c} \Sigma_- P_0 ^{\prime *}
 -\frac{1}{2} \Sigma_+ P_0^{\prime \prime *} + {\rm
 c.c.} .  
\end{eqnarray}
Using the definitions (\ref{app0}) -- (\ref{apps}) and defining
\begin{eqnarray}
 \label{5.11}
 w_0  &\equiv&  w (z_{2j}), \\
 w_1 &\equiv& \frac{i}{k} \left. \frac{\partial w}
 {\partial z} \right| _{z=z_{2j}}, \\
 \label{5.12}
 w_2 &\equiv& \left. \left( \frac{1}{k^2} 
 \frac{\partial ^{2}}{\partial z^{2}} - 2 \right) w \right| _{z=z_{2j}} , 
\end{eqnarray}
we  then arrive at the equations of motion
\begin{eqnarray}
 \label{5.15}
 \frac{\partial P_0}{\partial \tau} &=& -i \delta P_0 + w_0 \Sigma _+ ,\\
 \label{5.16}
 \frac{\partial P_{B}}{\partial \tau} &=& -i \delta P_{B} + 2 w_0 \Sigma _-
 -2 w_1 \Sigma_+ ,\\ 
 \label{5.17}
 \frac{\partial P_{C}}{\partial \tau} &=& -i \delta P_{C} +  w_2 \Sigma _+
 +2 w_1 \Sigma_-  , \\
 \label{5.18}
 \frac{\partial w_0}{\partial \tau} &=& -\frac{1}{2} \Sigma _+ P_0^* +{\rm
 c.c.}, \\
 \label{5.19}
 \frac{\partial w_1}{\partial \tau} &=& \frac{1}{2} \Sigma _- P_0^* 
 -\frac{1}{4} \Sigma _+ P_{B}^* -{\rm c.c.}, \\
 \label{5.20}
 \frac{\partial w_2}{\partial \tau} &=& \Sigma _+
 \left( P_0^* - \frac{1}{2} P_{C}^* \right) -
 \frac{1}{2} \Sigma _- P_{B}^*  +{\rm c.c.} .
\end{eqnarray}

Equations (\ref{4.17}), (\ref{4.18}) and (\ref{5.15}) --
(\ref{5.17}) with their complex conjugates and Eqs.~(\ref{5.18}) --
(\ref{5.20}) form a closed set of 13 equations for the
variables $\Sigma _+$, $\Sigma _-$, $P_0$, $P_{B}$, $P_{C}$ and their
complex conjugates and $w_0$, $w_1$ and $w_2$, i.e., 13 real variables
together. The equations are parametrized by 3 real parameters $\eta$,
$\delta$ and $\gamma$. Note that for $\gamma$ = 0 Eqs.
(\ref{4.17}), (\ref{4.18}), (\ref{5.15}) and (\ref{5.18}) are
identical to Eqs. (\ref{Sigma}), (\ref{driven}), (\ref{P}) and
(\ref{w}), respectively, with $P_{0}$ standing for $P$.  Even though
the number of equations and variables has now increased, they are
still relatively easy to solve numerically, and can realistically express
the properties of the system over rather long times.

The case of a {\em uniform active medium\/} embedded in a Bragg
grating calls either for a solution of the {\em full second-order
Maxwell equation,\/} without the spatial slow varying approximation,
or for a soluton of an {\em infinite set of coupled equations\/} for
$P_i$ and $w_i$ as in the case discussed in Ch.\ref{sec:colls}. This
makes the {\em present analysis principally at variance\/} with that
of \cite{Akoz98}, where the slow varying approximation for atomic
inversion and polarization is assumed, thus {\em arbitrarily
truncating\/} the infinite hierarchy of equations to its first two
orders.

\subsection{Energy densities}

To reveal the physical meaning of the quantities $\Sigma_{\pm}$ and $P$,
we express the energy density of the electromagnetic field as
\begin{equation}
W_F=(1/8)\hbar \omega_{c} \rho_{0} \left( |\Sigma _{+}|^2+|\Sigma _{-}|^2\right) ,
\label{efield}
\end{equation}
that of the TLS excitations (considering the limit of infinitely thin
TLS layers) as
\begin{equation}
W_A=(1/2)\hbar \omega _0\rho_{0} \left( 1-\sqrt{1-|P|^2}\right) ,  \label{eatom}
\end{equation}
and the  energy density of the TLS-field interaction as 
\begin{equation}
W_I=(1/2)\hbar \rho_{0} \tau _0^{-1}\ {\rm Im}\ \left( \Sigma _{+}P^{*}\right) .
\label{eint}
\end{equation}
{F}rom Eq. (\ref{efield}) we conclude that $|\Sigma _{+}|^2$ and
$|\Sigma _{-}|^2$ are proportional to the number of photons per TLS
(atom), in the standing-wave symmetric and anti-symmetric modes, whose
anti-nodes and nodes, respectively, coincide with the active layers
(see Fig.~\ref{figschem}). Since the interaction time $\tau _0$ (see
Eq. (\ref{tau0})) is usually much larger than the optical period $2\pi
/\omega_{c} $, the interaction energy is negligible in comparison with
the energies of the field and atomic excitations.


\subsection{Linearized spectrum}

To reach general understanding of the dynamics 
of the model, the first necessary step is to 
consider the spectrum produced by the linearized version of
Eqs. (\ref{driven}), (\ref{bef1}) and (\ref{bef2}), which describe {\em
weak fields\/} in the limit of infinitely thin TLS layers.  Setting
$w=-1$, and
\begin{subequations}
\begin{eqnarray}
\Sigma _{+}&=&Ae^{i(\kappa \zeta -\chi \tau )},
\\
\Sigma _{-}&=&Be^{i(\kappa
\zeta -\chi \tau )},
\\
P&=&Ce^{i(\kappa \zeta -\chi \tau )},  \label{eigen}
\end{eqnarray}
\end{subequations}
we obtain from the linearized equation (\ref{bef2}) that 
$C=i(\delta -\chi )^{-1}A$.
Substituting this into Eqs.(\ref{driven}) and (\ref{bef1}), we arrive at the
dispersion relation for the wavenumber $\kappa $ and frequency $\chi $ in
the form 
\begin{equation}
(\chi ^2-\kappa ^2-\eta ^2)(\chi -\delta )  
\times \left\{ (\chi -\delta )\left[ \chi ^2-\kappa ^2-(2+\eta ^2)\right]
+2(\eta -\delta )\right\} =0.  
\label{dispersion}
\end{equation}

\input #dispersion

Different branches of the dispersion relation generated by Eq. (\ref
{dispersion}) are shown in Fig.~\ref{fig1}.  The roots $\chi $ = $\pm
\sqrt{\kappa ^2+\eta ^2}$ (corresponding to the solid lines in
Fig.~\ref{fig1}) originate from the driven equation (\ref{driven}) and
represent the dispersion relation of the Bragg reflector with the gap
$|\chi |<\eta $ (cf. Eq.~(\ref {be3})), that does not feel the
interaction with the active layers. The degenerate root $\chi \equiv
\delta $ is trivial, as it corresponds to the eigenmode (\ref{eigen})
with $A=B=0$. Important roots are those given by the curly
brackets in Eq. (\ref{dispersion}) (shown by the dashed and
dash-dotted lines in Fig. \ref{figschem}), since they give rise to
nontrivial spectral features. They will be shown
below to correspond to bright or dark solitons in the indicated
(shaded) bands.

The frequencies corresponding to $k=0$ are 
\begin{equation}
\chi_0 =\eta \mbox{ and } \chi_{0,\pm}
=-\frac 12 \left( \eta -\delta \right) \pm \sqrt{2+\frac 14\left( \eta
+\delta \right) ^2},
\label{omega_0}
\end{equation}
while at $k^2\rightarrow \infty $ the asymptotic expressions for
different branches of the dispersion relation are $\chi =\pm k$ and $
\chi =\delta +2\left( \eta -\delta \right) k^{-2}$. Thus, the
linearized spectrum always splits into {\em two\/} gaps, separated by
an allowed band, except for the special case, $\eta =\eta _0\equiv
\frac 12\delta +\sqrt{1+\frac 14\delta ^2}$, when the upper gap closes
down. The upper and lower band edges are those of the periodic
structure, shifted by the induced TLS polarization in the limit of a
strong reflection. They approach the SIT spectral gap for forward- and
backward-propagating waves (\cite{Mant95}) in the limit of weak
reflection. The allowed middle band corresponds to a polaritonic 
(collective atomic polarization) excitation in the periodic structure.
It is different from single-photon hopping in a PBG via resonant 
dipole-dipole interactions (\cite{John95a}).


\section{Bright Solitons in RABR}
\label{S-bright}

\subsection{Standing (quiescent) self-localized pulses}
\label{S-bright-Q}

Stationary solutions of Eqs. (\ref{bef1}) and (\ref{bef2})
corresponding to bright solitons have been found by \cite{Kozh98a}.
Such solutions for the symmetric-mode field $\Sigma _{+}$ and
polarization $P$ are sought in the form
\begin{equation}
 \Sigma _{+}=e^{-i\chi \tau }{\cal S} (\zeta ),\qquad P=i\ e^{-i\chi \tau }
 {\cal P}(\zeta )
 \label{stationary}
\end{equation}
with real ${\cal P}$ and ${\cal S}$. Substituting this into
(\ref{bef2}), we eliminate ${\cal P}$ in favor of ${\cal S}$,
\begin{equation}
{\cal P}=-\frac{{\rm sign}(\chi -\delta )\cdot \ {\cal S} }{\sqrt{(\chi -\delta
)^2+|{\cal S} |^2}},  \label{psmall}
\end{equation}
and obtain an equation for ${\cal S} (\zeta )$, 
\begin{equation}
{\cal S} ^{\prime \prime }=(\eta ^2-\chi ^2){\cal S} -2{\cal S} \frac{(\eta -\chi
)\cdot {\rm sign}(\chi -\delta )}{\sqrt{(\chi -\delta )^2+{\cal S} ^2}},
\label{sigma''}
\end{equation}
where the prime stands for $d/d\zeta $. Equation (\ref{sigma''}) can
be cast into the form of the Newton's equation of motion for a particle
with the coordinate ${\cal S} (\zeta )$ moving in a potential $U({\cal
S})$:
\begin{equation}
{\cal S} ^{\prime \prime }=-U^{\prime }({\cal S} ),  \label{newt}
\end{equation}
where
\begin{equation}
U({\cal S} )=-\frac 12(\eta ^2-\chi ^2){\cal S} ^2
+2(\eta -\chi )
\cdot {\rm sign}
(\chi -\delta )\sqrt{(\chi -\delta )^2+{\cal S} ^2}.  \label{potential}
\end{equation}

\input #fpotentials

The potential gives rise to bright solitons (\cite{Newe92}), provided
it has two symmetric minima (see Fig.~\ref{fig-potential}).  As
follows from Eq. (\ref{potential}), the latter condition implies that
the quadratic part of the potential is concave, i.e., $|\chi |>\eta $,
and the second (asymptotically linear) part of the expression
(\ref{potential}) is convex, so that $\chi < \delta $.  Moreover, two
minima separated by a local maximum in ${\cal S}=0$ appear if $U''(0)
< 0$. From this inequality it follows that bright solitons can appear
in two frequency bands $\chi$, the lower band
\begin{equation}
\chi _1<\chi <{\rm min}\{\chi _2,-\eta ,\delta \},  \label{lower}
\end{equation}
and the upper band
\begin{equation}
{\rm max}\{\chi _1,\eta ,\delta \}<\chi <\chi _2 ,  \label{upper}
\end{equation}
where the boundary frequencies $\chi_{1,2}$ are given by
\begin{equation}
\chi _{1,2}
\equiv (1/2)\left[\delta - \eta \mp 
\sqrt{(\eta +\delta )^2+8} \right] .
\label{chi12}
\end{equation}
The lower band exists for all values $\eta >0$ and $\delta $, while
the upper one only exists for
\begin{eqnarray}
\delta >\eta -1/\eta ,
\label{eq:weak}
\end{eqnarray}
which follows from the requirement $\chi _2>\eta $ (see
Eq. (\ref{upper})).  On comparing these expressions with the spectrum
shown in Fig. \ref{fig1}, we conclude that part of the lower gap is
always empty from solitons, while the upper gap is completely filled
with stationary solitons in the weak-reflectivity case
(\ref{eq:weak}), and completely empty in the opposite limit. It is
relevant to mention that a partly empty gap has also been found in a
Bragg grating with second harmonic generation (\cite{Pesc97}), see Ch.
\ref{S-Boris}.

The bright soliton corresponds to the solution of the Newton equation
(\ref{newt}) with the `particle' sitting at time $-\infty$ on the
local maximum ${\cal S} = 0$, then swinging to one side and finally
returning to ${\cal S} = 0$ at time $+\infty$. Such solutions have
been found in an implicit form by \cite{Kozh98a}:
\begin{equation}
  {\cal S}(\zeta) = 2|\chi-\delta|{\cal R}(\zeta) \left( 1 - {\cal R}^{2}
  (\zeta) \right) ^{-1},  
\end{equation}
with
\begin{eqnarray}
  |\zeta| = \sqrt{2\left| \frac{\chi - \delta}{\chi - \eta}  \right|}
  \left[ (1-{\cal R}_{0}^{2})^{-1/2} \tan ^{-1}
  \sqrt{\frac{{\cal R}_{0}^{2}-{\cal R}^{2}}
  {1-{\cal R}_{0}^{2}}}  \right. \nonumber \\
  \left.
  + (2{\cal R}_{0})^{-1} \ln \left( \frac{{\cal R}_{0}
  + \sqrt{{\cal R}_{0}^{2}-{\cal R}^{2}}}{{\cal R}} \right)
  \right] ,
\label{zeta}
\end{eqnarray}
and
\begin{equation}
 {\cal R}_{0}^{2} = 1-\frac{|(\chi+\eta)(\chi-\delta)|}{2} 
\end{equation}
(note that ${\cal R}_0^2$ is positive under the above conditions
(\ref{lower})-(\ref{eq:weak})).  It can be checked that this
zero-velocity (ZV) gap soliton is always {\em single}-humped.  Its
amplitude can be found from Eq. (\ref{zeta}),
\begin{equation}
{\cal S}_{\max }=4 {\cal R}_0/\sqrt{\left| \chi +\eta \right| }.
\label{amplitude}
\end{equation}
The polarization amplitude ${\cal P}$ is determined by ${\cal S} $ via
Eq.~(\ref{psmall}).

To calculate the electric field in the antisymmetric
$\Sigma _{-}$ mode, we substitute 
\begin{equation}
 \Sigma _{-}=ie^{-i\chi \theta }{\cal A}
 (\zeta )
\end{equation}  
into Eq. (\ref{driven}) and obtain 
\begin{equation}
  {\cal A} ^{\prime \prime }+\left( \chi ^2-\eta ^2\right) {\cal A} =
  2{\cal P}^{\prime },
  \label{eqalpha}
\end{equation}
which can be easily solved by the Fourier transform, once ${\cal
P}(\zeta )$ is known. An example of bright solitons is depicted in
Fig.~\ref{fig-bright}.  Note that, depending on the parameters $\eta$,
$\delta$ and $\chi$, the main part of the soliton energy can be
carried either by the $\Sigma_{+}$ or the $\Sigma_{-}$ mode.

\input #mshape

The most drastic difference of these new solitons from the well-known
SIT pulses is that the area of the ZV soliton is not restricted to
$2\pi$, but, instead, may take an {\em arbitrary} value. As mentioned
above, this basic new result shows that the Bragg reflector can
enhance (by multiple reflections) the field coupling to the TLS, so as
to make the pulse area {\em effectively\/} equivalent to $2 \pi$. In
the limit of the small-amplitude and small-area solitons, ${\cal
R}_0^2 \ll 1$, Eq. (\ref{zeta}) can be easily inverted, the ZV soliton
becoming a broad {\rm sech}-like pulse:
\begin{equation}
{\cal S} \approx 2|\chi -\delta |{\cal R}_0\,{\rm sech}\left( \sqrt{2\left|
\frac{\chi -\eta }{\chi -\delta }\right| }{\cal R}_0\zeta \right) .
\label{small}
\end{equation}
In the opposite limit, $1-{\cal R}_0^2 \rightarrow 0$, i.e., for
vanishingly small $|\chi +\eta |$, the soliton's amplitude
(\ref{amplitude}) becomes very large, and further analysis reveals
that, in this case, the soliton is characterized by a {\em broad
central part\/} with a width $\sim \left( 1-{\cal R}_0^2\right) ^{-1/2}$
(Fig. \ref{fig-bright}(a)). Another special limit is $\chi -\eta
\rightarrow 0$. It can be checked that in this limit, the amplitude
(\ref{amplitude}) remains finite, but the {\em soliton width
diverges\/} as $|\chi -\eta |^{-1/2}$
(Fig. \ref{fig-bright}(b)). Thus, although the ZV soliton has a single
hump, its shape is, in general, strongly different from that of the
traditional nonlinear-Schr\"odinger (NLS) {\rm sech} pulse.

\subsubsection{Stability}

The stability of the ZV gap solitons was tested numerically, by means
of direct simulations of the full system (\ref{eq:system}), the
initial condition taken as the exact soliton with a small perturbation
added to it. Simulations at randomly chosen values of the parameters,
have invariably shown that the ZV GS are apparently {\em
stable}. However, the possibility of their dynamical and structural
instability needs be further investigated, as has been done in the
case of GS in a Kerr-nonlinear fiber with a grating by \cite{Bara98}
and \cite{Scho00}.

\subsection{Moving Solitons}
\label{S-bright-M}

Although the system of Eqs. (\ref{eq:system}) is not explicitly
Galilean- or Lorentz-invariant, translational invariance is expected
on physical grounds. Hence, a full family of soliton solutions should
have velocity as one of its parameters. This can be explicitly
demonstrated in the limit of the small-amplitude large-width solitons
[cf. Eq. (\ref{small})]. We search for the corresponding solutions in
the form $\Sigma _{+}(\zeta ,\tau )={\cal S} (\zeta ,\tau )\exp \left(
-i\chi_0\tau \right) ,\;P(\zeta ,\tau )=i {\cal P}(\zeta ,\tau
)\exp \left( -i\chi_0\tau \right) $ (cf. Eqs. (\ref{stationary})),
where $\chi_0$ is the frequency corresponding to $k=0$ on any of
the three branches of the dispersion relation (\ref{omega_0}) (see
Fig.  \ref{fig1}), and the functions ${\cal S} (\zeta ,\tau )$ and
${\cal P}(\zeta ,\tau )$ are assumed to be slowly varying in
comparison with $\exp \left( -i\chi_0\tau
\right)$. Under these assumptions, we arrive at the following
asymptotic equation for ${\cal S} (\zeta ,\tau )$:
\begin{eqnarray}
\left[ 2i\frac{\chi_0(\chi_0-\delta )^2-\eta +\delta }{(\chi
_0-\delta )^2} \frac{\partial}{\partial \tau } \right. 
&+&\frac{\partial^2}{\partial \zeta ^2}+ \label{NLS} \\ \left.
\frac{\chi_0-\eta }{(\chi_0-\delta)^3} 
|{\cal S}|^2 \right] {\cal S} &=& \left( \eta ^2-\chi
_0^2+2\frac{\chi_0-\eta }{\chi_0-\delta } \right) {\cal S} .
\nonumber 
\end{eqnarray}
Since this equation is of the NLS form, it has the full two-parameter
family of soliton solutions, including the moving ones (\cite{Newe92}).

In order to check the existence and stability of the moving
solitons numerically, the following procedure has been used by
\cite{Kozh98a}: Eqs. (\ref{eq:system}) were simulated for an
initial configuration in the form of the ZV soliton multiplied by
$\exp (ip\zeta )$ with some wavenumber $p$, in order to `push' the
soliton. The results demonstrate that, at sufficiently small $p$, the
`push' indeed produces a moving stable soliton
(Fig. \ref{fig:push}(a)). However, if $p$ is large enough, the
multiplication by $\exp (ip\zeta )$ turns out to be a more violent
perturbation, splitting the initial pulse into two solitons, one
quiescent and one moving (Fig. \ref{fig:push}(b)).

\input #msolit

Another one-parameter subfamily of moving GS was found in the exact
form of a phase-modulated $2 \pi$-soliton by \cite{Kozh95}:
\begin{equation}
\Sigma _{+}=A_0 \exp \left[ i\left( \kappa \zeta - \chi \tau \right) \right] 
{\rm sech}\left[ \beta \left( \zeta - v \tau \right) \right] ,
\label{moving}
\end{equation}
where $\chi$ is the detuning from the gap center, ${A}_0$ is the
amplitude of the solitary pulse, $\beta$ its width and $v$ its group
velocity.

Substituting $\partial_\tau {P}$ from Eq.(\ref{P}) into
Eq.(\ref{Sigma}), we may express $P$ in terms of $\Sigma_+$ and the
population inversion $w$.  Then, upon eliminating $P$ and using ansatz
Eq.(\ref{moving}), we can integrate Eq.(\ref{w}) for the population
inversion $w$, obtaining
\begin{equation}
\label{11}
w=-1 - \frac{ A^2_0 (\chi - \kappa /v )} {2 (\delta-\eta) }
\frac{1}{\cosh^2{\left[ \beta (\zeta - v \tau) \right] }}.
\end{equation}
Using these explicit expressions for $P$ and $w$ in Eqs.(\ref{Sigma})
and (\ref{P}), we reduce our system to a set of algebraic equations
for the coefficients $\kappa$, $\chi$ that determine the spatial and
temporal phase modulation, and the pulse width $\beta$ as functions of
the velocity $v$.
\begin{subequations} 
\label{12} 
\begin{eqnarray}
  2 (\chi - \kappa /v ) - ( 1 - 1/v^2) (\delta- \eta) &=&0, 
  \\ 
  (\chi - \delta)( \beta^2 v^2 - \beta^2 + \kappa^2 - \chi^2 +
  \eta^2 + 2) \qquad && \nonumber \\
  + 2 \beta^2 v^2 (\chi - \kappa /v) + 2 (\delta-\eta) &=& 0,
  \\ 
  ( \beta^2 v^2 - \beta^2 + \kappa^2 - \chi^2 + \eta^2 + 2) -
  2(\chi + \delta) (\chi - \kappa /v) &=& 0.
\end{eqnarray} 
\end{subequations}
The soliton amplitude is then found to satisfy $ |{A}_0| = 2 \beta v$,
exactly as in the case of usual SIT (see Ch. \ref{sec:3}). This
implies, by means of Eq.(\ref{moving}), that the area under the
$\Sigma_+$ envelope is $2 \pi$.

Let us consider the most illustrative case, when the atomic resonance
is exactly at the center of the optical gap, $\delta=0$. Then the
solutions for the above parameters are
\begin{subequations} 
\label{14} 
\begin{eqnarray}
  \kappa &=&- \frac{\eta}{2v} \frac{1-3v^2}{1-v^2}, \\ \chi &=&
  \frac{\eta}{2} \frac{1+v^2}{1-v^2}, \\ \beta^2 v^2&=& |A_0|^2/4 =
  \frac{8 v^2 (1-v^2) - \eta^2 (1+ v^2)^2}{4 (1-v^2)^2}.
\label{14'}
\end{eqnarray} 
\end{subequations}

In the frame moving with the group velocity of the pulse,
$\zeta'=\zeta-v \tau$, the temporal phase modulation will be $(\kappa
v - \chi)\tau $, which is found from Eq.(\ref{14}) to be equal to
$-\eta \tau$. Since $\eta$ is the (dimensionless) `bare' gap width
(see Ch. \ref{S:model}, this means that the frequency is detuned in the
moving frame exactly to the band-gap edge. The band-gap edge
corresponds (by definition) to a standing wave, whence this result
demonstrates that such a pulse is indeed a soliton, which does not
disperse in its group-velocity frame.

The allowed range of the solitary group velocities may be determined
from Eq.(\ref{14'}) through the condition $\beta^2>0$ for a given
$\eta$.  The same condition implies $ |\eta| < \eta_{\rm max} $, where
\begin{equation}
\label{kmax}
\eta^2_{\mbox{ \scriptsize max}} = \frac{8 v^2 (1-v^2)}{(1+v^2)^2}.
\end{equation}

It follows from Eq.(\ref{kmax}) that the condition for $2 \pi$ SIT gap
soliton (\ref{moving}) is $|\eta|<1$, $\eta_{\rm max}=1$
corresponding to $v=1/\sqrt{3}$.  This condition means that the
cooperative {\em absorption length} $c \tau_0 / n_0$ should be {\em
shorter than the reflection\/} (attenuation) {\em length\/} in the gap
$4c/(a_1 \omega_c n_0)$, i.e., that the incident light should be
absorbed by the TLS before it is reflected by the Bragg structure. In
addition, both these lengths should be much longer than the light
wavelength for the weak-reflection and slow-varying approximation to
be valid.

From Eq.(\ref{driven}) we find $\Sigma_- = \Sigma_+/v$. The envelopes
of both waves (forward and backward) propagate in the same direction;
therefore the group-velocity of the backward wave is in the direction
opposite to its phase-velocity! This is analogous to climbing a
descending escalator.

Analogously to Kerr-nonlinear gap solitons (Ch. \ref{S-Boris}), the
real part of the nonlinear polarization $\mbox{Re}P$ creates a
traveling `defect' in the periodic Bragg reflector structure which
allows the propagation at band-gap frequencies. The real part of the
nonlinear polarization is governed by the frequency detuning from the
TLS resonance. Exactly on resonance (which we here take to coincide
with the gap center) $\chi = \delta = 0$, $\mbox{Re}P=0$, and our
solutions (\ref{14}) yield imaginary values of the velocity $v$ and
modulation coefficient $\kappa$. The forward field envelope then
decays with the same exponent as in the absence of TLS in the
structure.  Because of this mechanism, SIT exists only on one side of
the band-gap center, depending on whether the TLS are in the region of
the higher or the lower linear refractive index. This result may be
understood as the addition of a near-resonant non-linear `refractive
index' to the modulated index of refraction of the gap structure. When
this addition compensates the linear modulation, soliton
propagation becomes possible (see Fig.\ref{figschem}).  On the `wrong' side
of the band-gap center, {\em soliton propagation is forbidden even in
the allowed zone}, because the nonlinear polarization then cannot
compensate even for a very weak loss of the forward field due to
reflection.

\input #fig2c

The soliton amplitude and velocity dependence on frequency detuning
from the gap center (which coincides with atomic resonance) are
illustrated in Fig.\ref{fig;4}. They demonstrate that forward soliton
propagation is allowed well within the gap, for $\chi$ satisfying $
(1-\sqrt{1-\eta^2})/\eta < \chi< (1+\sqrt{1-\eta^2})/\eta $.  In
addition to frequency detuning from resonance, the near-resonant GS
possesses another unique feature: spatial self-phase
modulation $\kappa \zeta$ of both the forward and backward field
components.

\subsection{Numerical Simulations}
\label{num_sim}

To check the stability of the analytical solution Eq.(\ref{moving}),
as well as the possibility to launch a moving GS by
the incident light field, numerical simulations of Eqs.(\ref{Maxw-P})
were performed by \cite{Kozh95}.  As the launching condition,
the incident wave was taken in the form ${\cal E}_{F} = A \exp{[i \chi
(t-t_0)]} / \cosh{[ \beta(t-t_0)/\tau_0]}$ without a backward wave
(${\cal E}_{B} = 0$) at the boundary of the sample $z=0$.  By varying
the detuning $\chi$ and amplitude $A$ we investigate the field
evolution inside the structure.  When these parameters are close to
those allowed by Eqs.(\ref{14}), (\ref{kmax}), we observe the formation
and lossless propagation of both forward and backward soliton-like
pulses with amplitude ratios predicted by our
solutions~(Fig.\ref{fig-3}).
By contrast, exponential decay of the forward pulse in the gap is
numerically obtained in the absence of TLS (Fig.\ref{fig-3.b}).

\input #gasol

\input #reflf


The analysis surveyed in Secs. \ref{S-bright-Q}-\ref{num_sim} 
strongly suggests, but does not
rigorously prove, that the solution subfamily (\ref{moving}) belongs
to a far more general two-parameter family, whose other particular
representatives are the exact ZV solitons (\ref{zeta}) and the
approximate small-amplitude solitons determined by Eq. (\ref{NLS}).

\subsection{Collisions between Gap Solitons}

An issue of obvious interest is that of collisions between GSs moving at
different velocities in RABR. In the asymptotic small-amplitude limit
reducing to the NLS equation (\ref{NLS}), the collision must be
elastic. To get a more general insight, we simulated collisions
between two solitons given by (\ref{moving}). The conclusion is that
the collision is {\em always inelastic\/}, directly attesting to the
nonintegrability of the model.  Typical results are displayed in
Fig. \ref{fig:4}, which demonstrates that the inelasticity may be
strong, depending on the parameters.

\input #collis



\section{Dark solitons in RABR}\label{S-dark}

\subsection[Existence conditions]{Existence conditions and the 
form of the soliton}

Dark solitons (DSs) in RABR have been studied by \cite{Opat99}.  They
are obtained similarly to the bright ones, by solving Eq.~(\ref{newt})
with the potential (\ref{potential}).  The potential will give rise to
DS's provided that it has two symmetric maxima (see
Fig.~\ref{fig-potential}). In this case the quadratic part of the
potential is convex, i.e., $|\chi |<\eta $, and the second
(asymptotically linear) part of the expression (\ref{potential}) is
concave, so that $\chi >\delta $.  {F}rom these two inequalities, a
simple necessary restriction on the model's parameters follows,
\begin{eqnarray}
\delta <\eta .  \label{podmin}
\end{eqnarray}
The condition for the existence of the
symmetric maxima determines the following frequency
interval $\chi $ (recall that $\eta $ is defined to be
positive): 
\begin{eqnarray}
 {\rm max}\{\delta ,-\eta \}<\chi <{\rm min}\{\chi _2,\eta \},
\label{interval}
\end{eqnarray}
Making use of (\ref{chi12}), one can easily check that, once the condition 
(\ref{podmin}) is satisfied, the DS-supporting band (\ref{interval}) 
{\em always} exists. The DS frequency range defined by Eq. (\ref{interval}) 
is marked by shading (to the right from zero) in Fig.~\ref{fig1}. 
The  maxima of the potential
are located at the points 
\begin{eqnarray}
 {\cal S} _M=\pm \sqrt{4(\eta +\chi )^{-2}-(\chi -\delta )^2},  \label{sigM}
\end{eqnarray}
which correspond to the polarization values  
\begin{eqnarray}
  {\cal P}_M=\mp \sqrt{1-(1/4)(\chi -\delta )^2(\chi +\eta )^2}.  \label{polar}
\end{eqnarray}
Integrating Eq. (\ref{newt}) by means of energy conservation in the
formal mechanical problem, we obtain ${\cal S} (\zeta )$ in an implicit form, 
\begin{eqnarray}
  \zeta  &=&\pm \frac 1{\sqrt{2}}\int_0^{\cal S} \frac{d{\cal S} _1}{\sqrt{%
  U_M-U({\cal S} _1)}}  \nonumber   \\
  &\equiv &\pm \frac 1{\sqrt{2}}\int_0^{\cal S} \frac{d{\cal S} _1}{\sqrt{U_M+%
  \alpha{\cal S} _1^2-\beta\sqrt{\gamma^2+{\cal S} _1^2}}},
  \label{implicit}
\end{eqnarray}
with
\begin{equation}
  \alpha \equiv \frac 12(\eta ^2-\chi ^2),\,\beta \equiv
  2(\eta -\chi ),\, \gamma \equiv \chi -\delta .
\end{equation}
The solution (\ref{implicit}) corresponds to a trajectory beginning at
the potential maximum $\pm{\cal S}_{M}$ at `time' $\zeta = - \infty $
and arriving at the other maximum, $\mp {\cal S}_{M}$ at `time' $\zeta
= \infty $ (see Fig.~\ref{fig-potential}). In terms of the $\Sigma
_{+}$ mode of the electric field, this is exactly a quiescent
(zero-velocity) DS with the background cw amplitude ${\cal S} _M$.

\input #darksolit

The integral (\ref{implicit}) can be formally expressed in terms of
incomplete elliptic integrals, but, practically, it is more helpful to
evaluate it numerically.  As in Ch. \ref{S-bright},
the polarization amplitude ${\cal P}$ is determined by 
${\cal S} $ via Eq.~(\ref{psmall}) and the 
amplitude of the $\Sigma_{-}$ mode is obtained by solving Eq.~(\ref{eqalpha}).
An example of the amplitude ${\cal S} $ for DS in the $\Sigma _{+}$
mode, together with the corresponding quantities ${\cal P}$ and 
${\cal A} $, are
plotted in Fig.~\ref{fig3}.

The energy density of the $|\Sigma _{+}|^{2}$ 
field mode always has the shape of a hole in
the background (see Fig.~\ref{fig4}). 
The energy density of $|\Sigma _{-}|^{2}$ has a hump, which
is the counterpart of the hole in the $\Sigma _{+}$ mode. The net
electromagnetic energy density may have either a hole 
(which never drops to zero) 
or a hump, depending on the system parameters $\eta $, $\delta $,
and the soliton frequency $\chi $.

\input #darkenerg


\subsection{The background stability}
\label{S-stability}

An obvious necessary condition for the stability of DS is the stability of
its cw background. To tackle this problem, we use Eq. (\ref{bef2}) to
eliminate $\Sigma _{+}$ in favor of $P$, 
\begin{eqnarray}
\Sigma _{+}=-\left( P_\tau -i\delta P\right) \left( 1-|P|^2\right) ^{-1/2},
\end{eqnarray}
and insert it into Eq. (\ref{bef1}). The resulting
equation for $P$ is linearized around the stationary value ${\cal P}_M$ 
(see Eq. (\ref{polar})), substituting 
\begin{eqnarray}
P={\cal P}_Me^{-i\chi \tau }\left[ 1+a(\zeta ,\tau )+i\ b(\zeta ,\tau )\right] ,
\label{line1}
\end{eqnarray}
where $a$ and $b$ are small real perturbations.  
We can look for its general solution in 
the form (cf. Eq. (\ref{eigen})), 
\begin{equation}
a(\zeta ,\tau )=a_0e^{i(\kappa\zeta -\Omega \tau )},
\; b(\zeta ,\tau )=b_0e^{i(\kappa\zeta
-\Omega \tau )},
\end{equation}
which leads to a dispersion relation for $\Omega $ and $\kappa$, that consists
of two parts: 
\begin{eqnarray}
  -(\chi\!-\!\delta)(\chi\!+\!\eta)^2 \Omega^3 - 
  2\left[ \chi (\chi\!-\!\delta)
  (\chi\!+\!\eta)^2 +2 \right]\Omega^2 
  \nonumber \\
  +\left[ (\eta\!+\!\chi)^3 (\eta \!- \!\delta)
  (\chi\!-\!\delta)\! -\! 8 \chi\! + \!(\chi\!-\!\delta)
  (\chi\!+\!\eta)^2 \kappa^2 \right]\Omega 
  \nonumber \\
  + (\eta^2 \!-\!\chi^2)\left[ 4-(\chi\!-\!\delta)^2 
  (\eta\!+\!\chi)^2 \right]
  + 4 \kappa^2 \ = \ 0 ,
  \label{stab1}
\end{eqnarray}
and
\begin{equation}
  -\Omega^3 - (3\chi\!-\!\delta)\Omega^2 
  +
  \left[ (\eta\!-\!\chi) (2\chi \!+\! \eta \!-\! \delta)\! +\! 
  \kappa^2 \right]\Omega 
  + (\chi\!-\!\delta) \kappa^2 \ = \ 0.
\label{stab2}
\end{equation}

\input #stabilit

As  checked numerically by \cite{Opat99}
for many values of $\eta $, $\delta $ and 
$\chi $ that support DS's according to the results obtained in the previous
section, all the roots of (\ref{stab1}) and (\ref{stab2}) are real for 
{\em  any} real $\kappa$. This implies the stability of the background for these
values of $\eta $, $\delta $ and $\chi $. 

Equations (\ref{stab1}) and (\ref{stab2}) represent new dispersion relations,
which are valid under the condition of strong background field ${\cal S}_{M}$
and replace the zero-field dispersion relations of Eq.~(\ref{dispersion}).
This kind of optical bistability can be compared to the distributed feedback
bistability with Kerr nonlinearity studied by~\cite{Winf79}.


\subsection{Direct numerical stability tests}

Even though there is no evidence of  background instability, 
it is necessary to
simulate the full system of the partial differential equations,
in order to directly
test the DS stability. 
Equations (\ref{bef1}) and 
(\ref{bef2}) have been integrated numerically by \cite{Opat99}, 
with initial conditions differing from the
exact DS solution by a
small perturbation added to it. 
The results strongly depend on the
parameters $\eta $, $\delta $ and 
$\chi $: for some values, an explosion of the initial
perturbation occurs, leading to a completely irregular pattern, 
whereas for others the DS shape remained {\em virtually undisturbed}. 
The dependence of the
stability on the parameters $\eta$ and $\chi$ with fixed $\delta$ is shown
in Fig.~\ref{figstab}. The darkest area of the DS parameter region 
corresponds to the stable regime where no instability has occurred
during the entire simulation time (typically, $\tau \sim$ 500). In the
rest of the parameter region, DS's are unstable: in the lightest
part of the DS region, the instability develops very quickly (at $\tau$ $<$
50), whereas in the intermediate part, the instability builds up relatively
slowly. As can be seen, the unstable behavior occurs closer to the 
boundaries of the existence region,
$\chi$ $=$  $\pm$ $\eta$ and $\chi$ $=$ $\delta$, whereas
along the boundary $\chi$ $=$ $\chi_{2}$ (corresponding to the DS-supporting 
background which degenerates into the trivial zero solution), DS's are stable.


\subsection{Coexistence of the dark and bright solutions}

A very interesting question is
whether the system can support bright and dark solitons
at the {\em same values\/} of the parameters. As mentioned above,  DS's
always exist in the frequency interval (\ref{interval}), once the inequality
(\ref{podmin}) is satisfied. On the other hand, bright solitons are
found in two frequency bands  $\chi $,
given in Eqs.~(\ref{lower}) and (\ref{upper}). From the discussion in
Ch.~\ref{S-bright}  it follows that
the DS frequency band {\em always coexists} with one or two bands
supporting the bright solitons. The
special case when there are {\em two}
bright-soliton bands coexisting with the DS band is singled out by the
condition 
\begin{eqnarray}
\eta -\frac 1\eta <\delta <\eta .  \label{special}
\end{eqnarray}
One can readily check that the coexisting frequency bands supporting
bright and dark solitons never overlap, i.e., quite naturally, the bright
and dark solitons cannot have the same frequency.


\subsection{Moving  dark solitons}
\label{S-moving}

Thus far we have considered only the quiescent DS's. A challenging question is
whether they also have their moving counterparts. Adding the velocity parameter
to the exact DS solution is not trivial, as the  underlying equations
(\ref{bef1}) and (\ref{bef2}) have no Galilean or Lorentzian invariance. The
physical reason for this is the existence of the special (laboratory) reference
frame, in which the Bragg grating is at rest. 

In principle it is possible, in analogy
to  the stationary solutions and Eq.~(\ref{stationary}), substitute functions
of the argument $(\zeta-v\tau)$ into the set (\ref{bef1}) 
and (\ref{bef2}) to obtain an ordinary differential equation. 
However, this would be a
complicated complex nonlinear equation of the third order, containing all the
lower-order derivatives, so that we would not be able to take 
advantage of the Newton-like
structure, as  in 
Ch.~\ref{S-bright} and \ref{S-dark}. Though it is possible to
solve such an equation numerically, it is
more suitable to deal with the original
set of partial differential equations,
in order to better understand the  nature of the
evolution.

In contrast to the case of bright solitons where the moving solutions
can be found by multiplying, in the initial conditions,
the quiescent DS by a factor proportional to exp$(i\kappa
\zeta)$ (see \cite{Kozh98a} and Ch.~\ref{S-bright}),
it has proven  possible to generate {\em stable moving\/} DS's from the 
quiescent ones in a different way (\cite{Opat99}).
To this end, recall that a DS corresponds to a transition between two
different values of the background cw field. 
The background field takes, generally, complex values 
(note the real values in the expressions (\ref{sigM}) and (\ref{polar}) are
only our choices adopted above for convenience). 
The quiescent DS corresponds to a transition
between two background values with phases differing by $\pi$. A principal
difference of the DS's in the present model from those in the NLS equation
(\cite{Kivs98}) is that here a moving DS is generated 
by introducing a {\em phase jump\/} $\neq \pi$ across the DS.

\input #movingdark

Thus, one can take the initial condition for the
system of  equations (\ref{bef1}) and (\ref{bef2}) as
\begin{eqnarray} 
 \label{initmov1}
 \Sigma_{+}(\zeta,0) &=& \cos \left( \frac{\phi}{2}\right)
 {\cal S}_{q}(\zeta) + i \sin  \left( \frac{\phi}{2}\right) {\cal S}_{M}, \\ 
 P &=&
 \cos \left( \frac{\phi}{2}\right) p_{q}(\zeta) + i \sin  \left(
 \frac{\phi}{2}\right) p_{M} ,
 \label{initmov2} 
\end{eqnarray} 
where ${\cal S}_{q}$ and ${\cal P}_{q}$ are
the (real) functions corresponding to the quiescent DS, 
${\cal S}_{M}$ and ${\cal P}_{M}$ 
are given by Eqs.~(\ref{sigM}), (\ref{polar})
and $\phi$ is the
deviation from $\pi$ of the background phase jump across DS. A typical result
obtained by means of this modification of the initial state
is displayed in Fig.~\ref{figmov}: the DS 
moves at a velocity that is
proportional to $\phi$. The resulting form of the 
moving DS is slightly different from that of the quiescent soliton. 
The moving DS appears to be stable over the entire simulation time.

\section[Light bullets]{Light bullets (spatiotemporal solitons)} 
\label{sec:lightbullets}

A promising direction is the study of solitons in resonantly absorbing
multi-dimensional (2D and 3D) media, in which the quasi-1D
periodic
structures can be realized as thin homogeneous layers set
perpendicular to
the direction of light propagation. In such media, {\em
spatiotemporal} solitons, i.e., those localized in all dimensions,
both transverse (spatial proper) and longitudinal (effectively,
temporal), may exist.  Spatiotemporal optical solitons or `light
bullets' (LBs) in various nonlinear media are surveyed in
Ch. \ref{S-intro}.

Here we are concerned with LBs in RABRs that 
consist of thin TLS layers embedded in a 2D- or 3D-
periodic dielectric medium. We will follow a recent analysis by
\cite{Blaa00a}, which extends an earlier prediction 
of stable LBs in uniform
2D and 3D SIT media (\cite{Blaa00}).

We start by considering a 2D SIT medium with a refractive index $n(z,x)$
periodically modulated in the propagation direction $z$, which represents
the quasi-one-dimensional Bragg grating. Light propagation in the medium is
described by the lossless Maxwell-Bloch equations (\cite{Newe92}): 
\begin{subequations}
\label{eq:SIT1}
\begin{eqnarray}
-i\frac{\partial^2 {\cal E}}{\partial x^2}+n^{2}
\frac{\partial {\cal E}}{\partial \tau }+
\frac{\partial {\cal E}}{\partial \zeta}+i\,(1-n^{2})\,{\cal E}-
{P} &=&0,  \label{eq:SIT11} \\
\frac{\partial P}{\partial \tau }-{\cal E} w &=&0,  \label{eq:SIT12} \\
\frac{\partial w}{\partial \tau }+
\frac{1}{2}({\cal E}^{\ast }{P}+{P}^{\ast }{\cal E})
&=&0.  \label{eq:SIT13}
\end{eqnarray}
\end{subequations}
Here (as in Sec. \ref{S:model-Maxwell}) 
${\cal E}$ and ${P}$ are the slowly varying amplitudes of the
electric field and medium's polarization, $w$ is the population
inversion, $\zeta$ and $x$ are longitudinal and transverse coordinates
(measured in units of the resonant-absorption length), and $\tau $ is
time (measured in units of the input pulse duration). The Fresnel
number, which governs the transverse diffraction in the 2D and 3D
propagation, was incorporated into $x$, and the detuning of the
carrier frequency $\omega _{0}$ from the central atomic-resonance
frequency was absorbed into ${\cal E}$ and $P$. To neglect the
polarization dephasing and inversion decay, we assume pulse durations
that are short on the time scale of the relaxation
processes. Equations (\ref{eq:SIT1}) are then compatible with the
local constraint $|P|^{2}+w^{2}=1$, which represents the so-called
Bloch-vector conservation. In a 1D case, i.e., in the absence of the
$x$-dependence and for $n(z,x)=1$, Eq.~(\ref{eq:SIT11}) reduces to the
sine-Gordon (SG) equation, which has a commonly known soliton
solution, see Eq.(\ref{2pi-SG}) and Ch.\ref{sec:3}.

To search for LBs in a 2D medium subject to a resonant periodic
longitudinal modulation, one may assume a periodic modulation of the
refractive index as per Eq.(\ref{be1}). The RABR is then
constructed by placing very thin layers (much thinner than $1/k_{c}$)
of two-level atoms, whose resonance frequency is close to $\omega
_{c}$, at {\it maxima} of this modulated refractive index.

The objective is to consider the propagation of an electromagnetic
wave with a frequency close to $\omega _{c}$ through a 2D RABR. Due to
the Bragg reflections, the electric field ${\cal E}$ gets decomposed
into forward- and backward-propagating components ${\cal E}_{F}$ and
${\cal E}_{B} $, which satisfy equations that are a straightforward
generalization of the 1D equations derived by \cite{Kozh95,Kozh98a},
and \cite{Opat99} (see also Eqs. (\ref{eq:max}) and (\ref{eq:bloch})
in this review):
\begin{subequations}
\label{eq:SIT4}
\begin{eqnarray}
-i \frac{\partial^3 \Sigma_+}{\partial \tau x^2} + 
i \frac{\partial^3 \Sigma_-}{\partial \zeta x^2} + 
\frac{\partial^2 \Sigma_+}{\partial \tau^2} - 
\frac{\partial^2 \Sigma_+}{\partial \zeta^2} \qquad &&  \label{eq:SIT41} \\
+\eta \frac{\partial^2 \Sigma_+}{\partial x^2} + \eta^{2}
\Sigma_{+}-2\frac{\partial P}{\partial \tau }-2i\eta {P}
&=&0,  \nonumber \\
-i\frac{\partial^3 \Sigma_-}{\partial \tau x^2} + i 
\frac{\partial^3 \Sigma_+}{\partial \zeta x^2} + 
\frac{\partial^2 \Sigma_-}{\partial \tau^2} -
\frac{\partial^2 \Sigma_-}{\partial \zeta^2} \qquad && \label{eq:SIT42} \\
-\eta \frac{\partial^2 \Sigma_-}{\partial x^2} + \eta^{2}
\Sigma_{-}+2\frac{\partial P}{\partial \zeta} &=&0, \nonumber \\
\frac{\partial P}{\partial \tau }+i \delta {P}-\Sigma_{+} w &=&0,  
\label{eq:SIT43}
\\
\frac{\partial w}{\partial \tau }
+\frac{1}{2}(\Sigma_{-}^{\ast} {P}+\Sigma_{+} {P}^{\ast })
&=&0.  \label{eq:SIT44}
\end{eqnarray}
\end{subequations}
Here $\Sigma_{\pm }$ is defined by Eq. (\ref{calesigma}), and
$\eta$ is a ratio of the resonant-absorption length in the
two-level medium to the Bragg-reflection length, which was
defined above by Eq. (\ref{eqeta}).

To construct an analytical approximation to the LB solutions, 
the starting point adopted by \cite{Blaa00a} is a subfamily of the 
exact 1D soliton solutions to Eqs.~(\ref{eq:SIT4}), which
was found by \cite{Kozh95} (see also Sec. 
\ref{S-bright-M} in this review) and is given by
Eqs.(\ref{moving}) and (\ref{11}).  
These solutions were taken with
parameters satisfying Eqs.(\ref{12}): 
\begin{subequations}
\label{m:params}
\begin{eqnarray}
A_0 &=& 2 \sqrt{\frac{\delta}{\eta} -1}, \quad 
\beta = \sqrt{\frac{\delta}{\eta} +1}, \quad 
v=-\sqrt{\frac{\delta - \eta}{\delta + \eta}}, \\
\kappa &=& -\sqrt{\delta^2 - \eta^2}, \quad {\rm and } \quad 
\chi=\delta.
\end{eqnarray}
\end{subequations}

These solutions were chosen as a pattern to construct an approximate
solution for LBs because the shape of the fields
$\Sigma_{+}$ and $\Sigma_{-}$ in the solutions is similar to that of
the SG soliton in the 1D uniform SIT medium (see Sec. \ref{sit_in_um}). 
Inspired by this analogy
and by the fact that there exist LBs in the uniform 2D SIT medium which
reduce to the SG solitons in the 1D limit (\cite{Blaa00}), one can search
for an approximate LB solution to the 2D equations~(\ref{eq:SIT4}),
which also reduces
to the exact soliton in 1D. To this end, the following
approximation was assumed:
\begin{subequations}
\label{eq:parm4}
\begin{eqnarray}
\Sigma_{+} &=&A_0 \sqrt{\mbox{\rm sech}\Theta _{1}\mbox{\rm
sech}\Theta _{2}} e^{i(\kappa \zeta - \chi \tau)+i\pi /4},
\label{eq:parm41} \\
\Sigma_{-} &=&\Sigma_{+}/v,
\label{eq:parm42} \\
{P} &=&\sqrt{\mbox{\rm sech}\Theta _{1}\mbox{\rm sech}\Theta _{2}}
\{(\tanh \Theta _{1}+\tanh \Theta _{2})^{2}+  \nonumber \\
&&\frac{\delta - \eta}{4\eta} C^{4} 
\bigl[ (\tanh \Theta _{1}-\tanh \Theta _{2})^{2}-2(
\mbox{\rm sech}^{2}\Theta _{1}+  \nonumber \\
&&\mbox{\rm sech}^{2}\Theta _{2})\bigr]^{2}\}^{1/2}
e^{i(\kappa \zeta - \chi \tau)+i\nu },\\
w&=&\left[ 1-|{P}|^{2}\right] ^{1/2},  \label{eq:parm44}
\end{eqnarray}
\end{subequations}
with $\Theta _{1}(\tau ,\zeta) \equiv \beta ( \zeta - v \tau ) +
\Theta_{0}+Cx$, $\Theta _{2}(\tau ,\zeta) \equiv \beta ( \zeta - v
\tau ) +\Theta _{0}-Cx$, the phase $\nu $ and coefficients $\Theta_0$ 
and $C$ being real constants, while the other parameters
are defined by Eqs.(\ref{m:params}).

The ansatz~(\ref{eq:parm4}) satisfies~Eqs. (\ref{eq:SIT41}) and
(\ref{eq:SIT42}) exactly, while Eqs. (\ref{eq:SIT44}) are satisfied to
order $\sqrt{\delta/\eta-1}C^{2}$, which requires that
$\sqrt{\delta/\eta-1} C^{2}\ll 1$. The ansatz applies for {\it
arbitrary} $\eta $, admitting {\em both} weak ($\eta \ll 1$) and
strong ($\eta >1$) reflectivities of the Bragg grating, provided that
the detuning remains small with respect to the gap frequency.
Comparison with numerical simulations of Eqs.~(\ref{eq:SIT4}), using
Eq. (\ref{eq:parm4}) as an initial configuration (a finite-difference
method, with Fourier transform scheme, described by \cite{Drum83}, was
used), tests this analytical approximation and shows that it is indeed
fairly close to a numerically exact solution; in particular, the shape
of the bullet remains within 98\% of its originally presumed shape
after having propagated a large distance, as is shown in
Fig.~\ref{BulletsFig1}.

\input #miriam

Three-dimensional LB solutions with axial symmetry have also been
constructed in an approximate analytical form and succesfully tested
in direct simulations, following a similar approach (\cite{Blaa00a}).
Generally, they are not drastically different from their 2D
counterparts described above.

A challenging problem which remains to be considered is the
construction of {\em spinning} light bullets in the 3D case
(doughnut-shaped solitons, with a hole in the center, carrying an
intrinsic angular momentum). Recently, spinning bullets were found by
means of a sophisticated version of the variational approximation in a
simpler 3D model, viz., the nonlinear Schr\"{o}dinger equation with
self-focusing cubic and self-defocusing quintic nonlinearities, by
\cite{Desy00}. Further direct simulations have demonstrated that these
spinning bullets (unlike their zero-spin counterparts) are always
subject to an azimuthal instability, that eventually splits them into
a few moving zero-spin solitons, although the instability can
sometimes be very weak (\cite{Miha00}).  At present, it is not known
whether spinning LBs can be completely stable in any 3D model.

It is relevant to stress that two- (and three-) dimensional LB solutions of
the variable-separated form, 
$\Sigma_{+}\sim \Sigma_{-}\sim f(\tau ,\zeta)\cdot
g(x)$, do {\em not} exist in the RABR model. 
Indeed, the substitution of this into Eqs.
(\ref{eq:SIT41}) and (\ref{eq:SIT42}) yields only a plane-wave solution of
the form $\Sigma_{\pm }\sim \exp \left( iA\tau +i B x \right) $, with
constant $A$ and $B$.

\section{Experimental prospects and conclusions}
\label{S-exper}

This review has focused on 
properties of solitons in RABR, combining a periodic refractive-index (Bragg)
grating and a periodic set of thin active layers (consisting of
two-level systems {\em resonantly\/} interacting with the field).  
It has been demonstrated that the RABR
supports a vast family of bright gap solitons, whose properties
differ substantially from their counterparts in periodic structures
with either cubic or quadratic {\em off-resonant\/} nonlinearies reviewed in
Ch.~\ref{S-Boris}.

The same RABR can support, depending on the initial conditions, either
dark or bright stable solitons, without any changes of the system
parameters, which is a unique feature for nonlinear optical media
(Ch.~\ref{S-bright}, \ref{S-dark}).  Zero-velocity dark solitons can
be found in an analytical form, as well as traveling dark solitons
with a constant phase difference ($\neq \pi$) of the background
amplitudes across the soliton (Ch.~\ref{S-dark}).  The latter property
is a major difference with respect to dark solitons of the NLS equation, whose
motion is supported by giving the background a nonzero wavenumber.

Depending on the values of the parameters, the frequency band of the
quiescent dark solitons coexists with one or two bands of the stable
bright ones, without an overlap.  Direct numerical simulations
demonstrate that some dark-soliton solutions are stable against
arbitrary small perturbations, whereas others are unstable when they
are close to the ``dangerous'' boundaries of their existence domain.

A multidimensional version of the RABR model, corresponding to a
periodic set of thin active layers placed at the maxima of the
refractive index, which is modulated along the propagation direction
of light has been considered too. It has been found to support stable
propagation of spatiotemporal solitons in the form of two- and
three-dimensional `light bullets' (LBs).

The best prospect of realizing a RABR which is adequate for observing
the solitons and light bullets discussed in Ch.~\ref{S-bright} -
Ch.~\ref{sec:lightbullets} is to use thin layers of {\em rare-earth
ions\/} (\cite{Grei99}) embedded in a spatially-periodic semiconductor
structure (\cite{Khit99}).  The two-level atoms in the layers should
be rare-earth-ions with the density of $10^{15}-10^{16}$\thinspace\
cm$^{-3}$, and large transition dipole moments.
The parameter $\eta $ can vary from 0 to 100 and the detuning is $\sim
10^{12}-10^{13}$\thinspace\ s$^{-1}$.

Cryogenic conditions in such structures can strongly extend 
the dephasing time $T_2$ and
thus the soliton's or LB's lifetime, well into the $\mu$sec range
(\cite{Grei99}), which would greatly facilitate the experiment.  The
construction of suitable structures constitutes a feasible
experimental challenge.

In a RABR with the transverse size of $10\,\mu $m, LBs can be
envisaged to be localized on the time and transverse-length scales,
respectively, $\sim 10^{-12}$ s and $1\,\mu $m.  The incident pulse
has uniform transverse intensity and the transverse diffraction is
strong enough. One needs $d^{2}/l_{\rm abs}\lambda _{0}<1$, where
$l_{{\rm abs}}$, $\lambda _{0}$ and $d$ are the resonant-absorption
length, carrier wavelength, and the pulse diameter, respectively 
(\cite{Slus74}). 
For $l_{{\rm abs}}\sim 10^{-3}$\thinspace\ m and $\lambda _{0}\sim
10^{-4}$\thinspace\ m, one thus requires $d<10^{-4}$ m, which implies
that the transverse medium size $L_{x}$ must be$\,$a few$%
\,\mu $m.

Effects of TLS dephasing and deexcitation in RABRs can be studied by
substituting the values $-i\delta - \Gamma_{2}\tau_{0}$ for the
frequency term $-i\delta$ in Eqs.~(\ref{5.15}) -- (\ref{5.17}) and the
loss terms $-\Gamma_{1}\tau_{0} (w_{0}+1)$ in Eq.~(\ref{5.18}),
$-\Gamma_{1} w_{1}$ in Eq.~(\ref{5.19}) and $-\Gamma_{1} (w_{2}-2)$ in
Eq.~(\ref{5.20}).  We have checked that these modifications {\em do
not} influence the qualitative behavior of the solutions on the time
scale $\tau \tau_{0}< 1/\Gamma_{1,2}$.

Let us now discuss the experimental conditions for the realization of
RABR solitons using {\em quantum wells\/} embedded in a semiconductor
structure with periodically alternating linear index of refraction
(\cite{Khit99}).  We can assume the following values: the average
refraction index is $n_{0} \approx 3.6$, the wavelength (in the
medium) $\lambda \approx 232$ nm, which corresponds to the angular
frequency $\omega_{c} \approx$ 2.26$\times 10^{15}$ s$^{-1}$.
Excitons in quantum wells can, under certain conditions (such as low
densities and proximity of the operating frequency to an excitonic
resonance, see \cite{Khit99}) may be regarded as effective two-level
systems (TLS's).  We consider their surface density to be $\approx
10^{10}$ -- $10^{11}$ cm$^{-2}$, which corresponds to a bulk density
$\rho_{0} \approx 10^{15}$ -- $10^{16}$ cm$^{-3}$. If we assume that
the excitons are formed by electrons and holes displaced by $\approx
1$ -- $10$ nm, then the characteristic absorption time $\tau_{0}$
defined in Eq.~(\ref{tau0}) is $\tau_{0} \approx 10^{-13}$ --
$10^{-12}$ s, and the corresponding absorption length is
$c\tau_{0}/n_{0}$ $\approx 10$ -- 100 $\mu$m.  The {\em dephasing
time\/} for excitons discussed by
\cite{Khit99} is $1/\Gamma_{2} \approx 10^{-13}$~s, which seems to be the
{\em chief limitation\/} 
of the soliton lifetime for this system. 
The structures shown in Figs.~\ref{fig-bright}, \ref{fig3}, \ref{fig4},
and \ref{figmov}, occupying regions of approximately 
100 absorption lengths
would require a device of the total width of approximately 1~mm to 1~cm, which
corresponds to $\approx$ $10^3$ to $10^4$ unit cells.
The modulation of the
refraction index can be as high as $a_{1} \approx 0.3$, so that the parameter
$\eta$ (see Eq.~(\ref{eqeta})) can vary from 0 to $100$. The
unit of the dimensionless detuning $\delta$ would represent a $10^{-3}$ --
$10^{-2}$ fraction of the carrier frequency. The intensities of the applied
laser field corresponding to $\Sigma_{\pm} \approx 1$ are then of the order
$10^{6}$ -- $10^{7}$ W/cm$^{2}$.  

In the work by \cite{Khit99}, the width of the active layers (quantum
wells) is considered to be 5 -- 20 nm, which corresponds to the
parameter $\gamma^{2}$ (see Eq.~(\ref{app-gam})) in the range
$10^{-3}$ -- $2\times 10^{-2}$. In the simulations discussed by
\cite{Opat99}, taking the largest of these values and the parameters
as in Fig.~\ref{fig3}, i.e., $\eta=0.6$, $\delta = -2$, and
$\chi=0.25$, we have observed the time evolution of the system
(\ref{4.17}) -- (\ref{5.20}).  As the initial condition, both the DS
solution corresponding to zero width of the active layers and the DS
solution including the finite width correction, have been taken.  In
both cases, the evolution was quite regular over the observed time
$\tau \approx 50$, and the zero-width solution (with the quantities
${\cal S, P, A}$ as given by Eqs.~(\ref{implicit}), (\ref{psmall}) and
(\ref{eqalpha})) started to change after $\tau \approx 10$.

We can now sum up the discussion of experimental perspectives for the 
realization of RABR solitons: {\it a}) The prospects appear to be good for 
gratings incorporating thin layers of rare-earth ions under cryogenic
conditions. {\it b}) The realization of these solitons in excitonic 
superlattices would require much longer dephasing times than those currently
achievable in such structures.


\section*{Acknowledgments}

We are grateful to M. Blaauboer for discussions and help.
G.K. acknowledges the support of ISF, Minerva and US-Israel BSF.
T.O. thanks the Deutsche Forschungsgemeinschaft for support. A.K.
acknowledges support of the Thomas B.\ Thriges Center for Quantum
Information.


\end{document}